\documentclass{jfm}
\usepackage{amsmath}
\usepackage{graphicx}
\usepackage{epstopdf, epsfig}
\usepackage{subcaption}
\usepackage{diagbox}
\usepackage{color, colortbl}
\usepackage{tikz}
\usetikzlibrary{arrows,positioning,decorations.pathreplacing} 

\tikzset{
    >=stealth',
    pil/.style={
           ->,
           thick,
           shorten <=2pt,
           shorten >=2pt,}
}

\graphicspath{ {images/} }
\definecolor{Gray}{gray}{0.9}

\shorttitle{Optimizing the control of transition to turbulence}
\shortauthor{A. Pershin, C. Beaume, T. S. Eaves and S. M.\ Tobias}

\title{Optimizing the control of transition to turbulence using a Bayesian method}

\author{
  Anton Pershin\aff{1,2}
  \corresp{\email{anton.pershin@physics.ox.ac.uk}},
  C\'edric Beaume\aff{2},
  Tom S. Eaves\aff{3}
  \and Steven M. Tobias\aff{2} 
}

\affiliation{
  \aff{1}Department of Physics, University of Oxford, Oxford OX1 3PU, UK
  \aff{2}School of Mathematics, University of Leeds, Leeds LS2 9JT, UK
  \aff{3}School of Science and Engineering, University of Dundee, Dundee DD1 4HN, UK
}

\begin{document}

\maketitle

\begin{abstract}

The nonlinear robustness of laminar plane Couette flow is considered under the action of in-phase spanwise wall oscillations by computing properties of the edge of chaos, i.e., the boundary of its basin of attraction.
Three measures are used to quantify the chosen control strategy on laminar-to-turbulent transition: the kinetic energy of edge states (local attractors on the edge of chaos), the form of the minimal seed (least energetic perturbation on the edge of chaos), and the laminarization probability (the probability that a random perturbation from the laminar flow of given kinetic energy will laminarize).
A novel Bayesian approach is introduced to enable the accurate computation of the laminarization probability at a fraction of the cost of previous methods.
While the edge state and the minimal seed provide useful information about the dynamics of transition to turbulence, neither measure is particularly useful to judge the effectiveness of the control strategy since they are not representative of the global geometry of the edge.  
In contrast, the laminarization probability provides global information about the edge and can be used to evaluate the control effectiveness by computing a laminarization score (the expected laminarization probability) and the associated expected dissipation rate of the controlled flow.
These two quantities allow for the determination of optimal control parameter values subject to desired constraints.
The results discussed in the paper are expected to be applied to a wide range of transitional flows and control strategies aimed at suppressing or triggering transition to turbulence.


\end{abstract}

\begin{keywords}
Authors should not enter keywords on the manuscript, as these must be chosen by the author during the online submission process and will then be added during the typesetting process (see http://journals.cambridge.org/data/\linebreak[3]relatedlink/jfm-\linebreak[3]keywords.pdf for the full list)
\end{keywords}

\section{Introduction}

\begin{figure}
        \centering
        \includegraphics[width=0.8\textwidth]{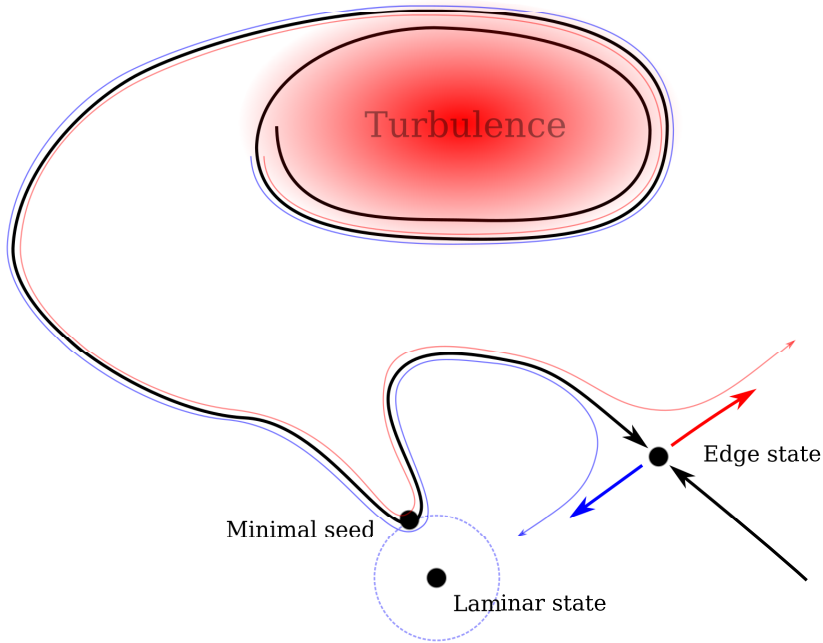}
        \caption{Sketch of phase space, illustrating the edge of chaos as a thick black line, turbulence, located by the shaded red region, and the laminar state as a fixed point. An attractor along the edge of chaos, called the {\it edge state} is represented together with its one-dimensional unstable manifold in thick colored arrows: the red side sends perturbations to turbulence while the right side sends them to the laminar flow. The minimal seed is indicated as well as, in dashed blue lines, the perimeter around the laminar flow on which perturbations have the same energy as the minimal seed. The thin lines show two trajectories starting from nearby positions: the red one eventually transitions to turbulence while the blue one eventually laminarizes.}
        \label{fig:edge_illustr}
    \end{figure}

Many shear flows feature a linearly stable laminar flow that is susceptible to transition to turbulence via a finite-amplitude instability \citep{Orszag1971, Romanov1973, Schmid2001, Meseguer2003, Barkley2016}.
Among these, plane Couette flow, i.e., the viscous flow between two parallel walls moving in opposite directions, is one of the most well studied examples.
In this configuration, both numerical and experimental studies confirm that turbulence can be sustained down to Reynolds number values as low as $325 \pm 10$ \citep{Dauchot1995, Duguet2010, Shi2013, Couliou2015}, which implies that a structure must exist in the phase space that separates decaying initial conditions from transitioning ones.
This structure, named the \textit{edge of chaos} \citep{Skufca2006}, is key to understanding the nonlinear route to turbulence and to designing control strategies for delay of  transition.
Figure \ref{fig:edge_illustr} shows a simplified representation of phase space to highlight some characteristics of the edge of chaos.

In addition to possessing a convoluted structure strongly interacting with turbulence \citep{Chantry2014}, 
the edge of chaos displays two objects that have received extensive attention in the literature: edge states and the minimal seed.
\emph{Edge states} are local attractors within the edge of chaos \citep{Schneider2008, Duguet2009, Schneider2010}.
These can be fixed points, periodic orbits or even chaotic sets, and are characterized by the fact that they only have one unstable direction.
A growing body of research suggests that edge states act as important mediators during the laminar-to-turbulent transition \citep{Cherubini2011, Khapko2016, Kreilos2016}.
It is thus tempting to choose the edge state energy as an indicator of the nonlinear robustness of the laminar flow, and to monitor its value when controlling the flow to either reduce or enhance the robustness of the laminar flow.
Such a perspective was considered by \cite{Rinaldi2018}, who found that the edge state energy may either increase or decrease (i.e. the related fixed point may shift away from or towards the laminar fixed point) in response to the introduction of non-uniform viscosity in plane channel flow.
As a consequence, they suggested that the basin of attraction of the laminar flow may grow or shrink, making the laminar flow either more or less robust to finite-amplitude perturbations.
The impact of control strategies on edge states is not guaranteed however.
The addition of polymer to Poiseuille flow, for example, does not impact the properties of its edge states \citep{Xi2012}. Further, the edge state and its energy are only a local measures of the edge of chaos and do not inherently contain information about the global structure of the edge or the likelihood of transition.

On the other hand, a direct measure of the nonlinear robustness of the laminar flow is provided by the \textit{minimal seed}, the initial condition along the edge of chaos that is the closest energetically to the laminar flow \citep{Pringle2012, Cherubini2013, Kerswell2018}. 
Energetically lower initial conditions are all contained within the basin of attraction of the laminar flow while there exists energetically higher initial conditions that trigger turbulence \citep[for a detailed discussion, see][]{Kerswell2018}.
There is emerging evidence that the minimal seed represents the most likely transition scenario, as it may be related to an instanton trajectory in the large deviation theory of noisy systems \citep{Lecoanet2018}.
Motivated by the fact that the minimal seed is the smallest-energy perturbation from the laminar flow that can trigger turbulence, \cite{Rabin2014} tuned control parameters in wall-oscillated plane Couette flow to maximize its kinetic energy as a way to improve the robustness of the laminar flow. 
However, minimal seeds have a very specific spatial structures which allows them to most efficiently trigger turbulence, and as such it is likely that the structure of the edge nearby to the minimal seed is somewhat cusped, with the minimal seed lying at the end of a narrow intrusion of the turbulent side of the edge into the basin of attraction of the laminar flow. Because of this, it is unlikely that changes in the energy of the minimal seed will have a significant impact on the overall size and shape of the basin of attraction of the laminar flow, and hence its robustness to a generic perturbation.

To circumvent the problems that arise when considering only local properties of the edge, an alternative perspective on the assessment of the robustness of the laminar flow has recently been proposed.
Instead of focusing on particular invariant objects of the edge of chaos, it characterizes the basin of attraction of the laminar flow globally via the \textit{laminarization probability}, i.e., the probability that a random perturbation of the laminar flow decays as a function of its kinetic energy \citep{Pershin2020}. 
In a sense, the laminarization probability measures the relative volume of the basin of attraction of the laminar state, thereby providing information on the global structure of the edge of chaos rather than looking at its local features.
\citet{Pershin2020} successfully used it to quantify the nonlinear stability of laminar plane Couette flow at several values of the Reynolds number and under the action of spanwise wall oscillations. However, the calculation of the laminarisation probability in \citet{Pershin2020} required an extremely large number of simulations to be performed to achieve statistically converged results. 
Here, we introduce a novel Bayesian approach to calculating the laminarisation probability that requires substantially fewer simulations to provide useful results.
As such, a significantly larger set of control parameters in plane Couette flow under the action of spanwise wall oscillations may be considered, in order to assess the perfomance of this particular turbulence control measure.

In this paper, we compute edge states, minimal seeds and laminarization probabilities to study the nonlinear robustness of plane Couette flow in the presence of spanwise wall oscillation and determine the conditions that minimize the flow sensitivity to finite-amplitude perturbations.
In doing so, we shed light on the connection between these flow features and the assessment of control strategies.
The next section is devoted to the set-up of the problem. 
Section 3 discusses the use of the edge state and minimal seed energies as quantifiers of the control efficiency.
Section 4 provides a detailed explanation and verification of the new Bayesian procedure used to estimate the laminarization probability, followed in section 5 by the results obtained by the application of this procedure.
These results are augmented by the introduction of two scalar criteria that can be used to quantify the efficiency of control strategies: the laminarization score, which represents the probability that a perturbation drawn from a given energy distribution laminarizes; and the expected dissipation rate, which quantifies the expected energy required to produce the controlled flow.
The paper concludes with a discussion in section 6.

\section{Plane Couette flow under spanwise wall oscillations}

We consider plane Couette flow, i.e., the flow driven by two infinite plates separated by a gap $2h$ and moving in opposite directions at speed $U$.
We subject the flow to sinusoidal in-phase wall oscillations in the spanwise direction with amplitude $UW_{osc}$, frequency $U \omega / h$ and phase $\phi$.
After non-dimensionalization, the Navier--Stokes equation and the incompressibility condition read:
\begin{eqnarray}
  \label{navsto}
&\partial_t \boldsymbol{u} + (\boldsymbol{U} \cdot \nabla) \boldsymbol{U} = -\nabla p + \displaystyle\frac{1}{\Rey} \nabla^2 \boldsymbol{u} + \left( \displaystyle\frac{1}{\Rey} \partial_{yy} W - \partial_{t} W \right) \boldsymbol{e}_z,\\
  \label{incomp}
&\nabla \cdot \boldsymbol{u} = 0,
\end{eqnarray}
where $\Rey = U h/\nu$ is the Reynolds number, $\nu$ is the kinematic viscosity, $p$ is the pressure and $\boldsymbol{e_z}$ is the unit vector in the spanwise direction.
In writing these equations, we have used the usual decomposition of the full flow field $\boldsymbol{U}$ into the laminar solution $\boldsymbol{U_{\text{lam}}} = [y, 0, W(y, t)]$ and the incompressible perturbation, or turbulent velocity, $\boldsymbol{u}$ such that $\boldsymbol{U} = \boldsymbol{u} + \boldsymbol{U_{\text{lam}}}$.
The presence of spanwise wall oscillations modifies the laminar solution of plane Couette flow by adding a time-dependent spanwise component $W(y, t)$:
\begin{eqnarray}
W(y, t) = &\displaystyle\frac{W_{osc}}{\cosh 2\Omega + \cos 2\Omega} \Big[ \left[ \cosh\Omega y_+ \cos\Omega y_- + \cosh\Omega y_- \cos\Omega y_+ \right] \sin(\omega t + \phi) \nonumber \\ 
& + \left[ \sinh\Omega y_+ \sin\Omega y_- + \sinh\Omega y_- \sin\Omega y_+ \right] \cos(\omega t + \phi) \Big],
\end{eqnarray}
where $\Omega = \sqrt{\omega \Rey/2}$ and $y_{\pm} = y \pm 1$.
More details can be found in the work of \cite{Rabin2014} and in Appendix \ref{app:deriv_osc_laminar}.
This decomposition  allows the laminar solution to absorb the (time-dependent) no-slip boundary condition $\boldsymbol{U}(x, \pm 1, z, t) = [\pm 1, 0, W_{osc} \sin(\omega t + \phi)]$ in such a way that the incompressible perturbation satisfies homogeneous boundary conditions in $y$.
We consider a periodic domain in the streamwise (period $\Gamma_x = 4 \pi h$) and in the spanwise (period $\Gamma_z = 32 \pi h /15$) directions and fix the Reynolds number to $\Rey = 500$ hereafter, a value significantly larger than that necessary to sustain turbulence: $\Rey > \Rey_c = 325 \pm 10$ \citep{Shi2013, Dauchot1995}.
These basic flow conditions are identical to those in \cite{Pershin2020}.

We use a suitably modified version of \textit{Channelflow} \citep{Gibson2014channelflow} to  numerically integrate equations (\ref{navsto}--\ref{incomp}) for the perturbation $\boldsymbol{u}$.
The streamwise and spanwise directions are discretized using $N_x=32$ and $N_z=34$ Fourier coefficients and the wall-normal direction is discretized using $N_y=33$ Chebyshev coefficients \citep{Pershin2020}.
A third-order semi-implicit backward differentiation scheme with time step $\triangle t = 1 / \Rey$ is used to advance the flow in time.

\begin{figure}
    \includegraphics[width=1.0\textwidth]{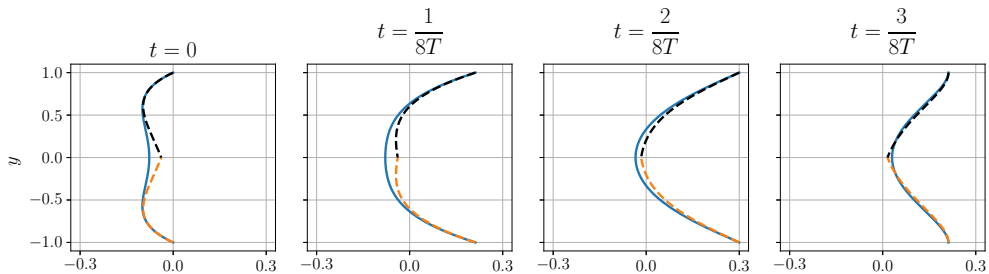}
    \caption{Time-evolution of the spanwise component of the laminar flow (blue curves) in the presence of in-phase wall oscillations for $W_{osc} = 0.3$, $\omega = 1/64$ and $\Rey = 500$ represented at $4$ different phases of the oscillation, where $T = 2\pi / \omega$ is the oscillation period. The orange and black dashed curves denote the Stokes boundary layers associated with the bottom and top walls respectively and defined by the following expressions: $W(y, t) =~W_{osc} e^{-\Omega \tilde y} \sin\left(\omega t - \Omega \tilde y\right)$, where $\Omega = \sqrt{\omega \Rey /2}$, $\tilde y = y + 1$ for the bottom wall and $\tilde y = 1 - y$ for the top wall. For these parameter values, the depth of penetration is $\delta \approx 0.5$.}
    \label{fig:in_phase_lam_flow}
\end{figure}

Modifying the plane Couette flow configuration by adding wall oscillations creates Stokes boundary layers consisting of transverse waves emanating from the top and bottom walls and travelling a distance $\delta = 1/\Omega$ toward the centre of the domain.
For a small enough depth of penetration ($\delta \lesssim 0.5$) or, equivalently, large enough frequency ($\omega \gtrsim 1/64$ for $\Rey = 500$), the spanwise flow can be thought of as a combination of two such boundary layers, as illustrated in figure \ref{fig:in_phase_lam_flow} for an oscillation amplitude $W_{osc} = 0.3$.

It is therefore reasonable to expect that the effect of the oscillating walls will propagate a depth $\delta$ into the domain and so the interior of the flow will not be affected by wall oscillations when the frequency is too large, so that efficient control via in-phase spanwise wall oscillations requires $\omega \le 1$.
Indeed, $\omega = 1$ corresponds to negligible depths of penetration: $\delta \approx 0.06$ for $Re = 500$.
On the other hand, the wall oscillation period should be at most comparable to the typical time it takes for an initial condition to transition to turbulence, which is $O(10^2)$ in this domain \citep{Pershin2020}, corresponding to $\omega \ge 1/128$.
We thus consider $\omega \in [1/128; 1]$, together with $W_{osc} \in [0; 0.5]$.

From an engineering standpoint, the addition of spanwise oscillations to plane Couette flow requires increased energetic input to counter viscous dissipation and maintain the flow.
Following ideas expressed by \cite{Rabin2014}, the necessary energy corresponds to the dissipation rate:
\begin{equation}
\label{eq:diss_rate}
\epsilon(W_{osc}, \omega) = \displaystyle\frac{\omega}{4 k \pi \Gamma_x \Gamma_z \Rey} \int \limits_{0}^{2 k \pi / \omega} \int \limits_{0}^{\Gamma_x} \int \limits_{-1}^{1} \int \limits_{0}^{\Gamma_z} |\nabla \times (\boldsymbol{U_{lam}} + \boldsymbol{u})|^2 \mathrm{d}z \, \mathrm{d}y \, \mathrm{d}x \, \mathrm{d}t,
\end{equation}
where time-averaging is performed over $k$ periods of wall oscillations depending on the length of the available time series.
For the laminar flow, this formula reduces to
\begin{equation}
\label{eq:lam_diss_rate}
\epsilon_{lam}(W_{osc}, \omega) = \displaystyle\frac{1}{\Rey} \left[ 1 + \frac{W_{osc}^2 \Omega}{2} \cdot \frac{\sinh(2 \Omega) - \sin(2 \Omega)}{\cosh(2 \Omega) + \cos(2 \Omega)} \right].
\end{equation}
In the absence of control, the dissipation rate for the laminar flow is $\epsilon_{lam}(W_{osc}=0) = 1/\Rey$.
For $\Rey = 500$ and within the considered ranges for $W_{osc}$ and $\omega$, the dissipation rates can be well approximated by:
\begin{equation}
\epsilon_{lam}(W_{osc}, \omega) = \displaystyle\frac{1}{\Rey} \left( 1 + \frac{W_{osc}^2 \sqrt{\omega \Rey}}{2\sqrt{2}} \right).
\end{equation}
Lastly, the dissipation rate that is calculated according to equation (\ref{eq:diss_rate}) for a turbulent flow realization will be referred to as $\epsilon_{turb}(W_{osc}, \omega)$.

\begin{figure}
    \includegraphics[width=1.0\textwidth]{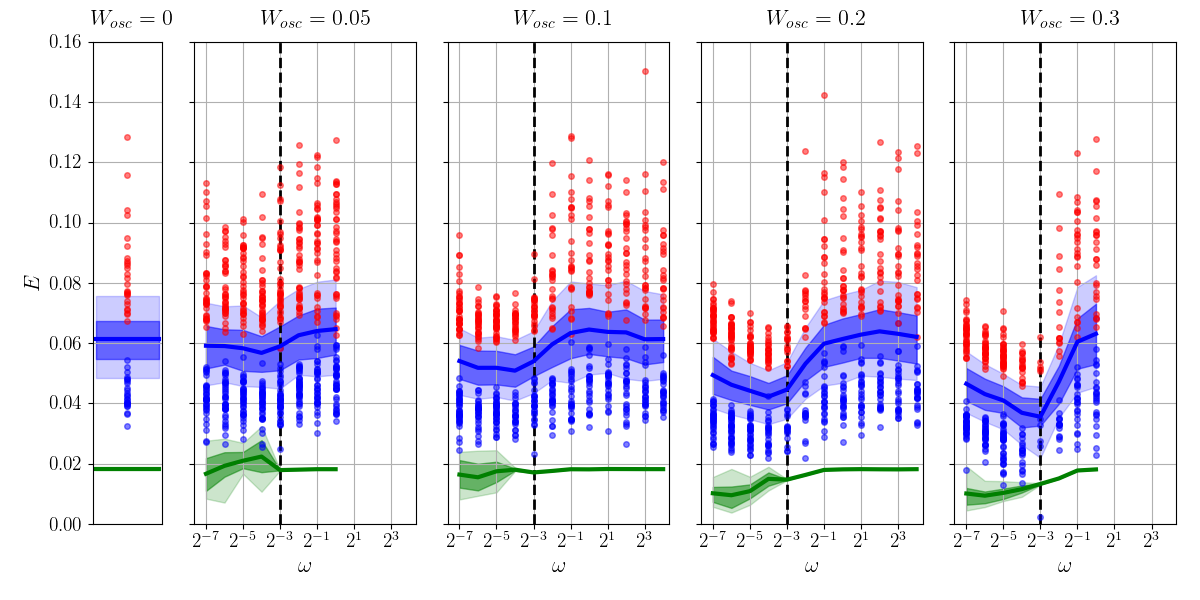}
    \caption{Distribution of the turbulent kinetic energy $E$ (blue) and the edge state kinetic energy (green) plotted as a function of frequency $\omega$ for amplitudes ranging from $W_{osc} = 0$ to $W_{osc} = 0.3$. Red and blue dots denote the block maxima and minima as explained in the text. Thick lines, dark bands and light bands denote the mean, interquartile range and interdecile range of the distributions respectively. The vertical dashed line corresponds to the reference frequency $\omega = 1/8$.}
    \label{fig:turb_attractor_estimate}
\end{figure}

To further characterize the flow, we also define the turbulent kinetic energy:
\begin{equation}
\label{eq:E_def}
E = \frac{1}{2}\langle\boldsymbol{u}, \boldsymbol{u} \rangle = \frac{1}{2}||\boldsymbol{u}||^2 = \frac{1}{4 \Gamma_x \Gamma_z}\int \limits_{0}^{\Gamma_x} \int \limits_{-1}^{1} \int \limits_{0}^{\Gamma_z} \boldsymbol{u} \cdot \boldsymbol{u} \; \mathrm{d}z \, \mathrm{d}y \, \mathrm{d}x.
\end{equation}
The mean, interquartile and interdecile ranges of the turbulent kinetic energy are shown in figure \ref{fig:turb_attractor_estimate} for amplitude values ranging from $W_{osc} = 0.05$ to $W_{osc} = 0.3$.
For $\omega \geq 1$, the mean turbulent kinetic energy does not vary much with $\omega$ and is close to that observed for the uncontrolled case $W_{osc} = 0$.
Reducing the frequency of the forcing below $\omega = 1/2$ leads to the decrease of the mean turbulent kinetic energy, which reaches a minimum for $\omega = 1/16$, except for $W_{osc} = 0.3$, where the minimal value of the mean turbulent kinetic energy is obtained for $\omega = 1/8$.
Additionally, increasing the oscillation amplitude leads to more rapid turbulence decay and we did not obtain conclusive data for $W_{osc} > 0.3$.
As we approach the minimizing forcing frequency, the kinetic energy distribution becomes more compact, as shown by the shrinking interquartile and interdecile ranges and by the reduced spread of the block extrema obtained by dividing the time series into non-overlapping blocks of equal length ($500$ time units in Figure \ref{fig:turb_attractor_estimate}) and collecting global extrema for each of these blocks.
This effect can barely be observed for $W_{osc} = 0.05$ but becomes more pronounced as the amplitude is increased.

\begin{figure}
    \includegraphics[width=1.0\textwidth]{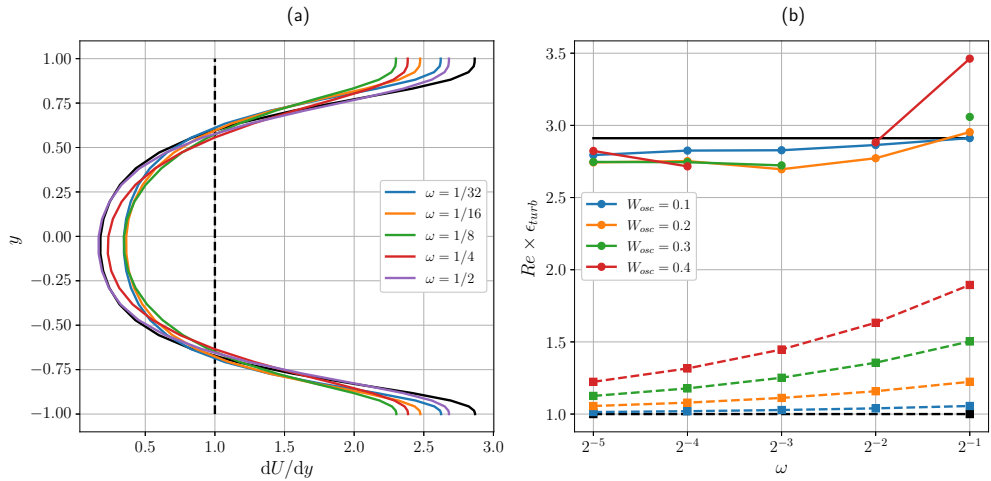}
    \caption{(a) Time-averaged wall-normal shear rate associated with the streamwise velocity in the presence of wall oscillations for $W_{osc} = 0.3$ and plotted for various frequencies $\omega$. (b) Laminar (dashed) and turbulent (solid) dissipation rates plotted as a function of frequency $\omega$ for amplitudes varying from $W_{osc} = 0.1$ to $W_{osc} = 0.4$. The results for $W_{osc} = 0.5$ are omitted because, at this amplitude, it was virtually impossible to sustain turbulence for the necessary time duration to reliably compute the turbulent dissipation rate. In both plots, dashed and solid black lines correspond to laminar and turbulent flows in the uncontrolled case ($W_{osc} = 0$) respectively.}
    \label{fig:turb_prof_and_diss}
\end{figure}

The decrease of the mean turbulent kinetic energy as the frequency of the control is increased towards the aforementioned optimum is accompanied with a reduced variation across the domain of the averaged shear rate.
Figure \ref{fig:turb_prof_and_diss}(a) shows how the shear stress of the time-averaged streamwise turbulent velocity profile changes as a function of the frequency for $W_{osc} = 0.3$.
The wall-normal shear rate variations across the domain are minimal for $\omega = 1/8$, the same value that minimizes the turbulent kinetic energy (see figure \ref{fig:turb_attractor_estimate}).

Finally, the dependence of the turbulent dissipation rate $\epsilon_{turb}(W_{osc}, \omega)$ on the control parameters is shown in Figure \ref{fig:turb_prof_and_diss}(b) and highlights the fact that in general either decreasing the wall oscillation frequency or increasing the oscillation amplitude leads to a decrease in the turbulent dissipation rate.

\section{Scalar measures of the robustness of the laminar flow}

\subsection{Edge states}
\label{sec:edge_states}

The edge of chaos is the manifold separating decaying initial conditions from those transitioning to turbulence.
Complete knowledge of this manifold would allow for the determination of the relative state space volume occupied by the basin of attraction of the laminar flow.
Characterizing the laminar flow robustness can thus be achieved via the exhaustive characterization of the edge of chaos.
Unfortunately, the edge has a highly complex structure \citep{Skufca2006, Schneider2007, Chantry2014}, and full characterization is virtually impossible to achieve.
To bypass this difficulty, the edge manifold is usually characterized by more accessible measures such as edge states, which are local attractors on the edge of chaos, and the minimal seed, which is the point on the edge of chaos that is closest to the laminar fixed point.

\begin{figure}
    \begin{subfigure}{0.49\textwidth}
        \centering
        \includegraphics[width=1.0\textwidth]{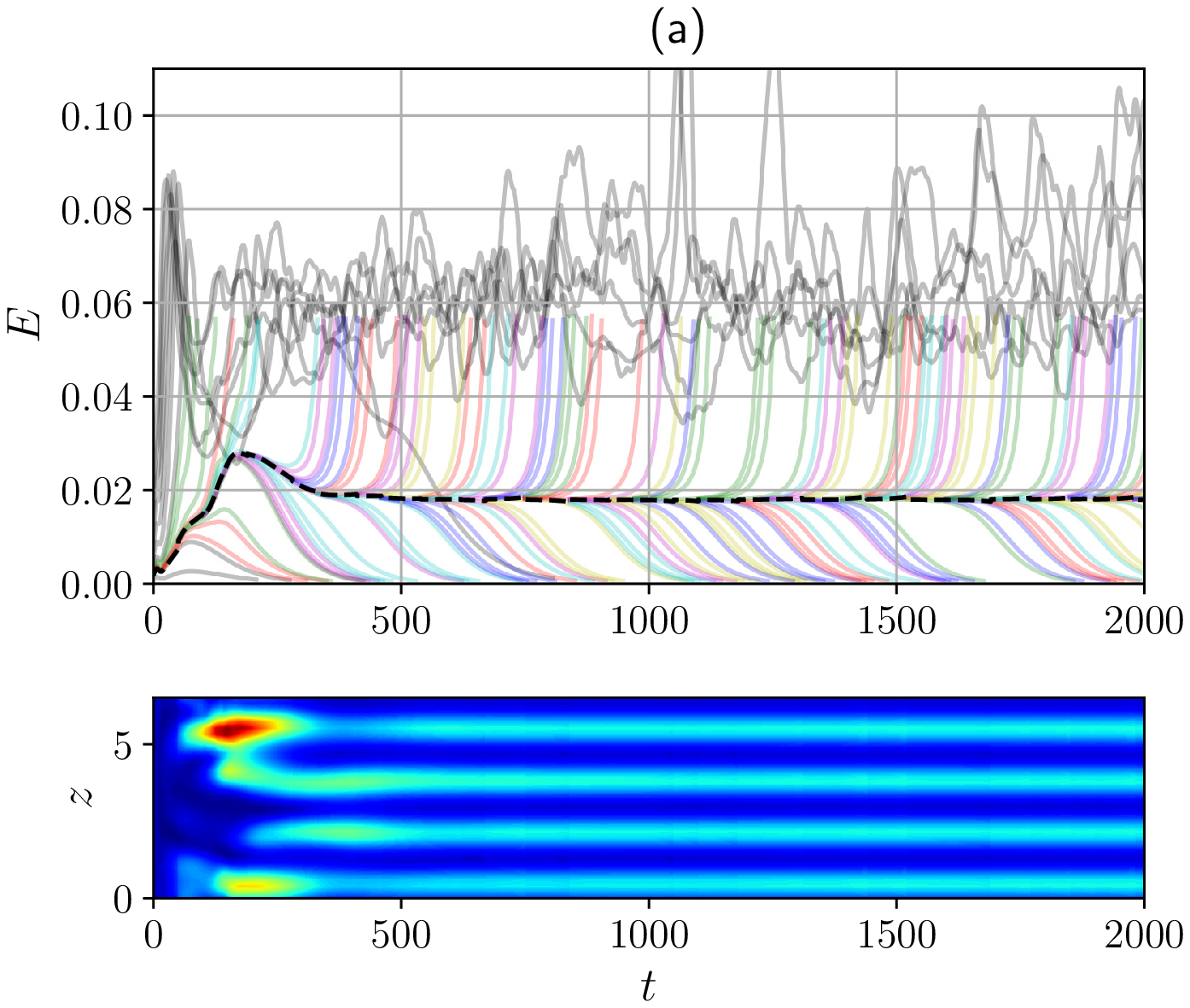}
    \end{subfigure}
    \begin{subfigure}{0.49\textwidth}
        \centering
        \includegraphics[width=1.0\textwidth]{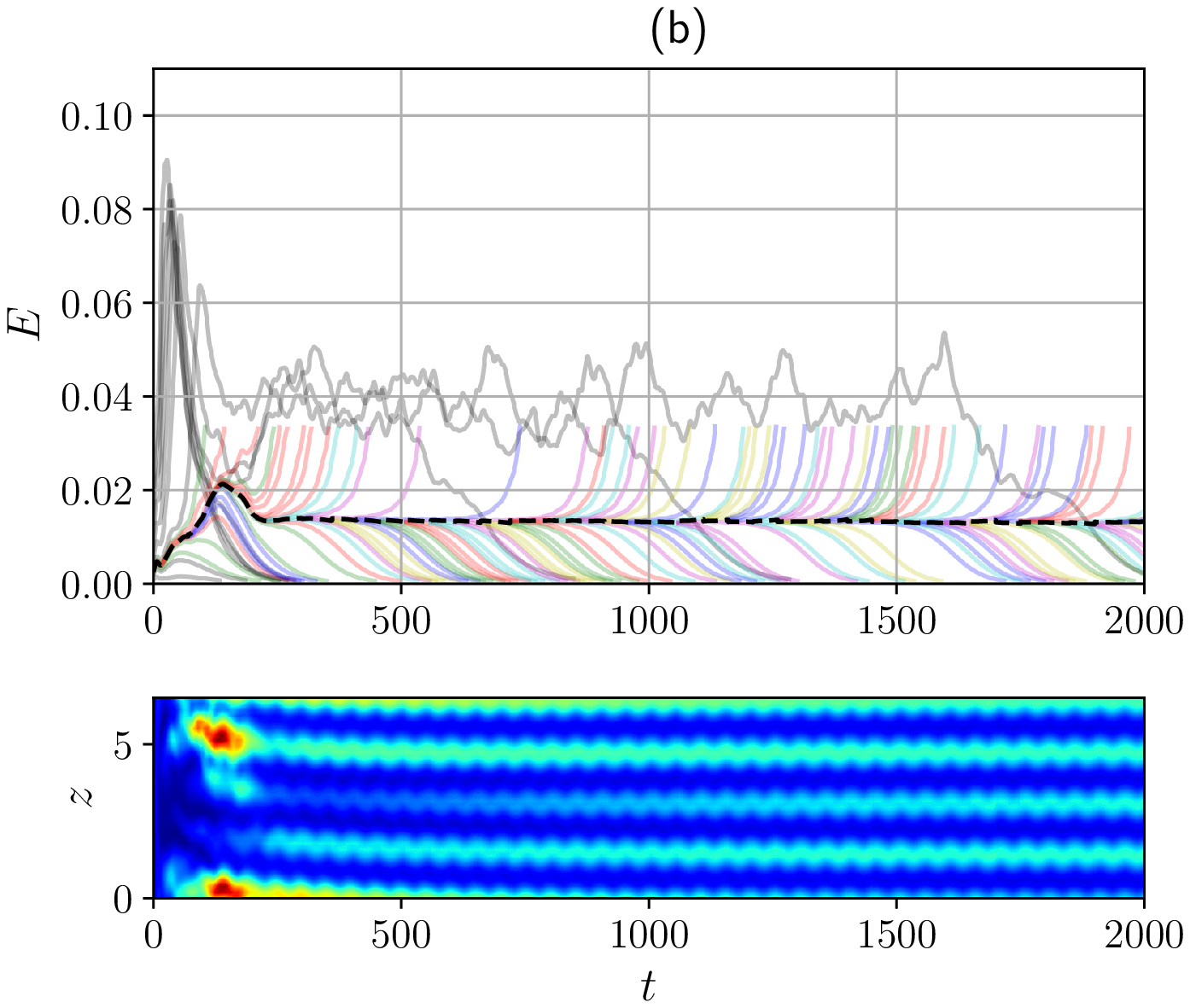}
    \end{subfigure}
    \begin{subfigure}{0.49\textwidth}
        \centering
        \includegraphics[width=1.0\textwidth]{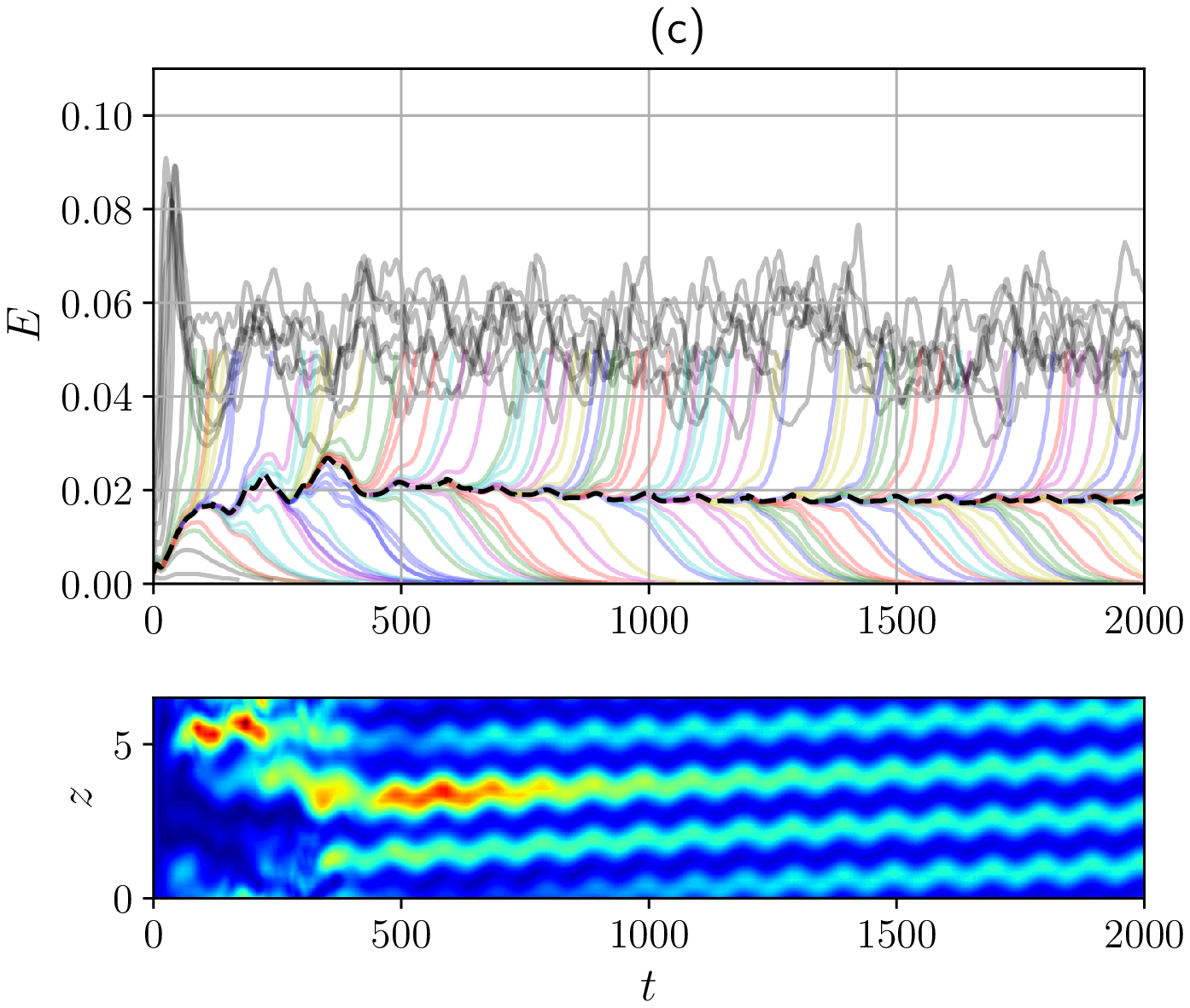}
    \end{subfigure}
    \begin{subfigure}{0.49\textwidth}
        \centering
        \includegraphics[width=1.0\textwidth]{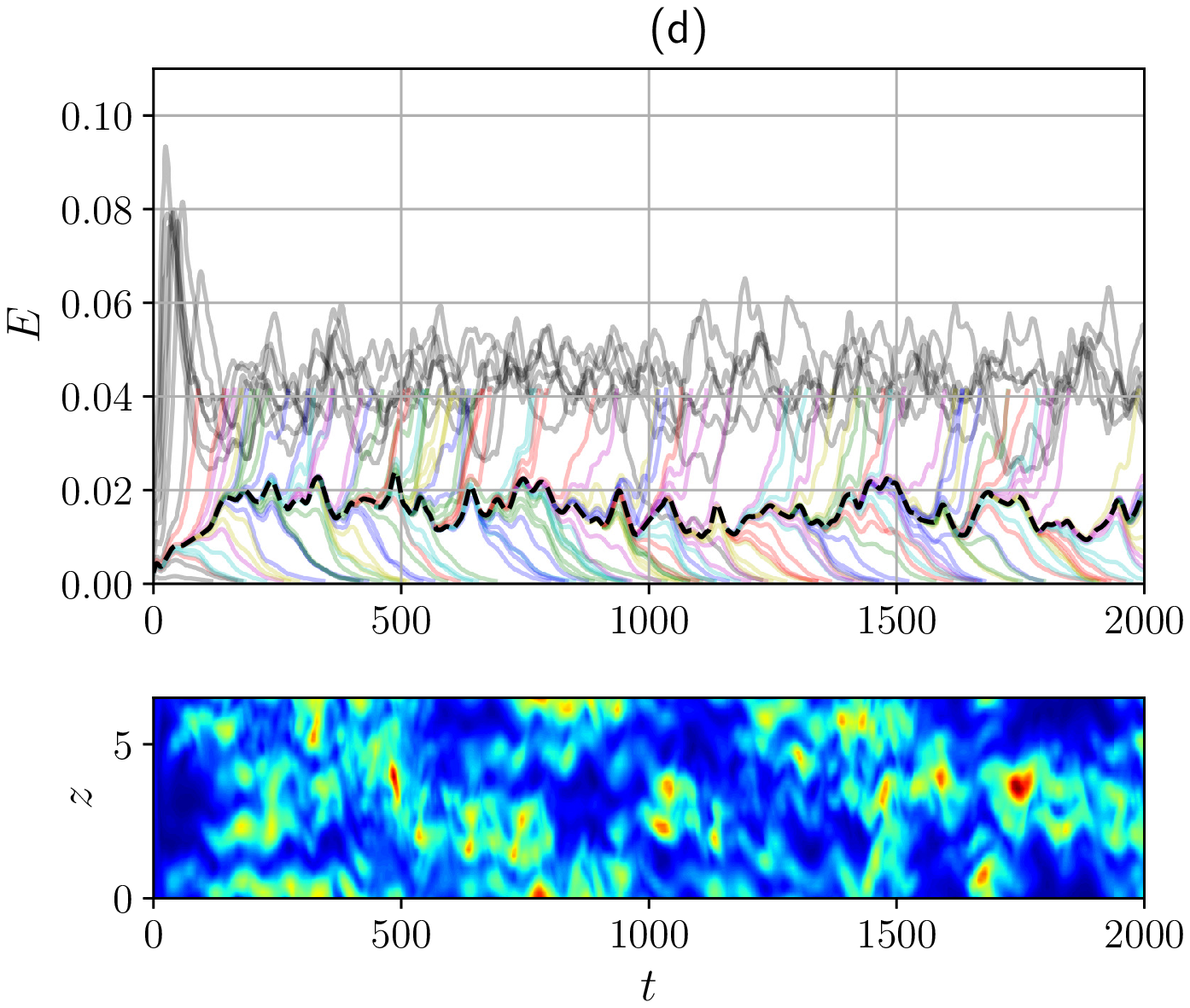}
    \end{subfigure}
    \caption{Visualisation of the edge tracking procedure for (a) $W_{osc} = 0.4, \omega = 1$, (b) $W_{osc} = 0.3, \omega = 1/8$, (c) $W_{osc} = 0.1, \omega = 1/16$ and (d) $W_{osc} = 0.2, \omega = 1/16$. In each panel, the top plot shows the time-evolution of the kinetic energy of the perturbation from the laminar flow used to get the edge tracking started (gray), of trajectories associated with subsequent iterations of the algorithm (colors) and of the trajectory along the edge (dashed black curve). The bottom plot of each panel shows the time-evolution 
 of the $xy$-averaged kinetic energy of the perturbation associated with the edge trajectories. The edge trajectories are found to approach edge states that can be classified according to the dynamics of the perturbation from the laminar flow: equilibrium (a), periodic orbit (b), relative periodic orbit (c) and chaotic trajectory (d), depending on the amplitude and frequency of wall oscillations (see diagram \ref{fig:edge_energy}(a) for details). The periodic orbit in (b) and the relative periodic orbit in (c) bear clear structural resemblance with the steady edge states obtained for larger control frequencies (a) which, in turn, are virtually indistinguishable from the edge state in the uncontrolled system.}
    \label{fig:edgetracking_examples}
\end{figure}

\begin{figure}
    \includegraphics[width=1.0\textwidth]{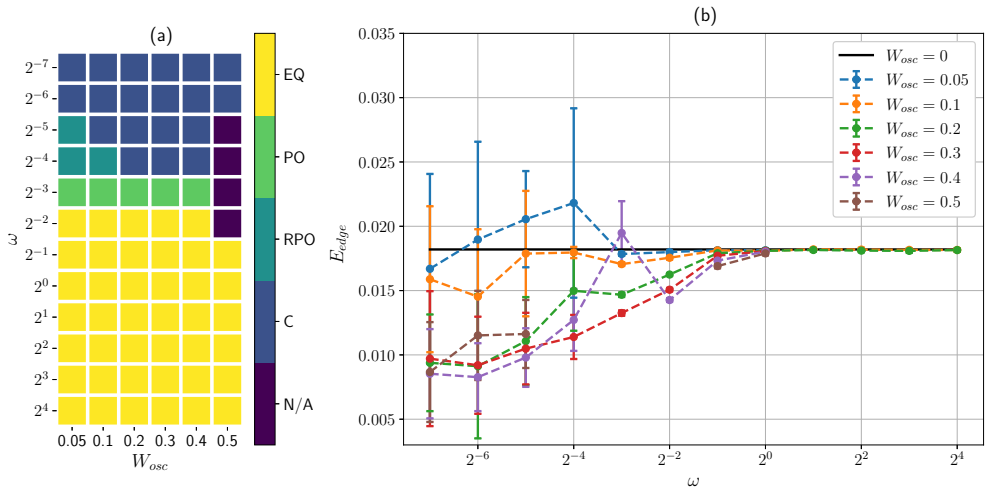}
    \caption{(a) Classification of the edge states as equilibria (EQ), periodic orbits (PO), relative periodic orbits (RPO) and chaotic trajectories (C) as a function of $W_{osc}$ and $\omega$. (b) Dependence of the mean edge state energy $E_{edge}$ and associated standard deviation (error bars) on the amplitude $W_{osc}$ and frequency $\omega$ of the wall oscillations. The edge state results are omitted for $W_{osc} = 0.5, \omega = 1/4, 1/8, 1/16, 1/32$ in both (a) and (b) since the control strategy almost completely suppresses turbulence at these parameter values.}
    \label{fig:edge_energy}
\end{figure}

We compute the edge states for control parameter values $W_{osc} \in [0.05; 0.5]$ and $\omega  \in [1/128; 1]$ using \textit{edge tracking} \citep{Skufca2006}.
Typical examples of the resulting edge trajectories, represented by the turbulent kinetic energy, are shown in figure \ref{fig:edgetracking_examples}.
The edge states  can be classified according to their dynamics.
We find that edge states can take the form of equilibria, periodic orbits, relative periodic orbits or chaotic objects depending on the control amplitude and frequency.
We present this classification in figure \ref{fig:edge_energy}(a) as a function of the control parameter values.
For large enough control frequencies, the edge state invariably takes the form of a steady state, virtually identical to that in the uncontrolled system (see figure \ref{fig:edgetracking_examples}(a)).
Decreasing the frequency adds temporal variability to the edge state, turning it into a periodic orbit (see figure \ref{fig:edgetracking_examples}(b)) and finally into a chaotic object (see figure \ref{fig:edgetracking_examples}(d)).
For the smallest amplitudes considered, $W_{osc} = 0.05$ and $W_{osc} = 0.1$, we also observed that the edge state could take the form of relative periodic orbits for frequencies between those associated with periodic orbits and chaotic edge states (see figure \ref{fig:edgetracking_examples}(c)).
We expect that more relative periodic edge states can be identified with a more refined discretization of the control frequency.
Transitions between the identified equilibria, periodic orbits and relative periodic orbits may be continuous as a function of $W_{osc}$ and $\omega$, implying that all equilibria are actually relative periodic orbits with nearly vanishing spanwise group velocity, or result from bifurcations: further refinement of the parameter space landscape to uncover a precise scenario is beyond the scope of this work.
Decreasing $\omega$ further yields steady edge states as expected from the asymptotic regime produced when $\omega \to 0$.

To further characterize the dependence of the edge states on the control parameters, we build the distribution of the edge state kinetic energy by taking the edge trajectories for the time-interval $t \in [1000; 1000 + 2 n \pi / \omega]$, where $n$ denotes the number of periods of wall oscillations taken into account ($n=2$ for the smallest frequency $\omega=1/128$ and $n \geq 4$ for larger frequencies; larger values of $n$ do not significantly modify the results).
The resulting dependence of the mean edge state energy and its standard deviation on the amplitude and frequency of the wall oscillations are shown in figure \ref{fig:edge_energy}(b).
Wall oscillation decreases the mean edge state energy in the vast majority of the cases.
Contrary to the turbulent kinetic energy (see figure \ref{fig:turb_attractor_estimate}), the edge state energy tends to decrease as the frequency of the forcing is decreased even when $\omega<1/8$, for most of the amplitude values.
One can however observe similarities between the behaviour of the kinetic energy of turbulent flows and that of the edge state under control: increasing the control amplitude tends to lower the kinetic energy associated with both distributions implying that both objects (the turbulent saddle and the edge state) are closer to the laminar state.
We also further note that the control frequency at which the edge state becomes unsteady is close to that associated with the minimum of the averaged turbulent kinetic energy.

\subsection{Minimal seeds}

The minimal seed is the initial condition of smallest turbulent kinetic energy $E=E_c$ which is on the edge manifold separating the basins of attraction of the laminar and turbulent states.
In practice, it is impossible to find the minimal seed exactly, but instead approximations to it which lie on either side of the edge manifold may be computed using techniques derived from nonlinear nonmodal stability theory \citep{Kerswell2018}. 
This involves iteratively solving a nonlinear energy optimisation problem which yields solutions that are only guaranteed to be \emph{locally} optimal in terms of the closest approach of the edge manifold.
However, minimal seeds found in plane Couette flow using a variety of different approaches, and for a variety of parameter values, all appear to share the same qualitative initial structure and evolve according to similar dynamics \citep{Monokrousos2011,Rabin2012,Duguet2013,Cherubini2013,Eaves2015}.
This lends confidence that these local minima may in fact be global minima, or that they at least represent dynamically significant initial conditions. The minimal seed found here for $W_{osc}=0$ is also similar to these other plane Couette flow minimal seeds.

For ease of discussion, we refer to the approximation which lies on the turbulent side of the edge manifold as \emph{the minimal seed}, and its turbulent kinetic energy is an upper bound on $E_c$. 
The turbulent kinetic energy of the approximation which lies on the laminar side of the edge manifold provides a lower bound on the (locally optimal) value of $E_c$; for the converged results presented below, the difference in turbulent kinetic energy between the two approximations that bracket $E_c$ is $2.5\times 10^{-8}$, or at most 0.2\% of $E_c$. Unfortunately, two of the results below did \emph{not} converge; it is well-known that the minimal seed optimisation problem struggles to converge owing to the sensitive dependence on initial conditions associated with the turbulent attractor \citep{Pringle2012,Rabin2012,Kerswell2018}, as do related optimisation problems in turbulence control \citep[see \textit{e.g.}][]{Vishnampet2015}.
The convergence issue is compounded further if the edge state itself is also a chaotic set \cite[see][]{Eaves2015}, and the two non-converged results in this work do indeed have chaotic edge states (c.f. figure \ref{fig:edge_energy}(a)).

For $W_{osc}\neq 0$, we perform a restricted search for the minimal seed, fixing the phase of the wall oscillation to be $\phi=0$.
A family of initial conditions may be found by searching for minimal seeds at different values of $\phi$, and the initial condition with the smallest turbulent kinetic energy over all values of $\phi$ is the minimal seed for this system. By fixing $\phi=0$, we instead find an upper bound for the minimum turbulent kinetic energy in this family.
However, \citet{Rabin2014} found that, for $\Rey = 1000$, the minimal seed with $\phi=0$ is almost optimal across all values of $\phi$, and that turbulent kinetic energies for different values of $\phi$ do not vary much.

\begin{figure}
  \centering 
	\includegraphics[width=0.55\textwidth]{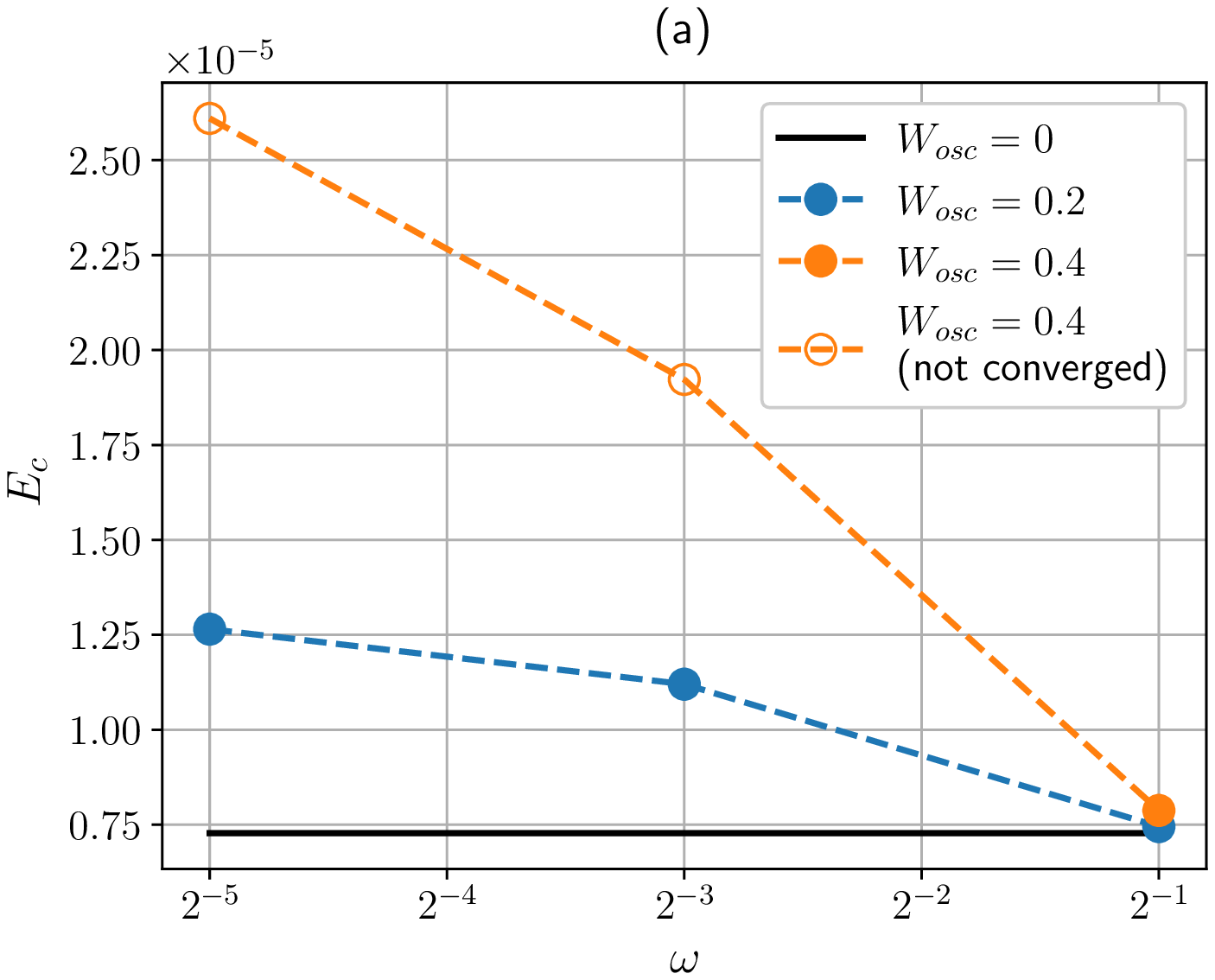}
	\begin{minipage}{0.9\textwidth}
	\begin{center}
	\includegraphics[width=0.45\textwidth]{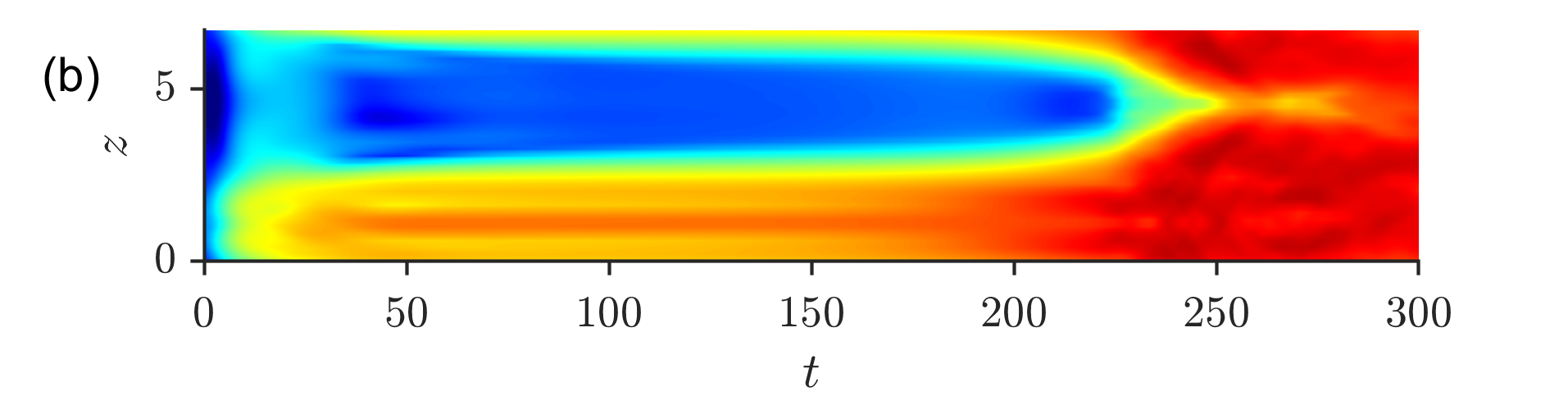} \\	    
	\end{center}
	\includegraphics[width=0.45\textwidth]{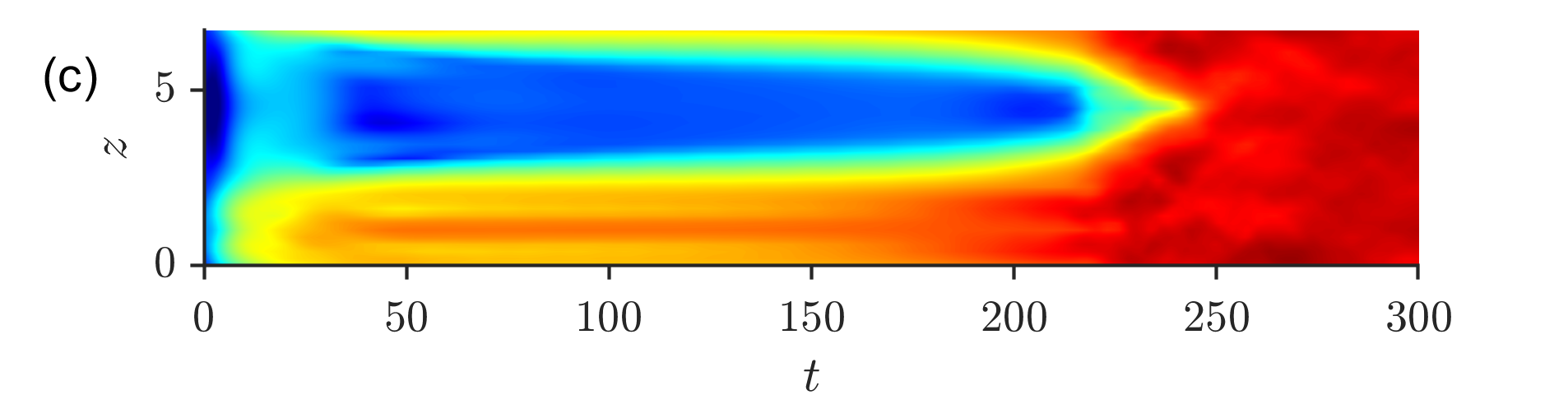}
	\includegraphics[width=0.45\textwidth]{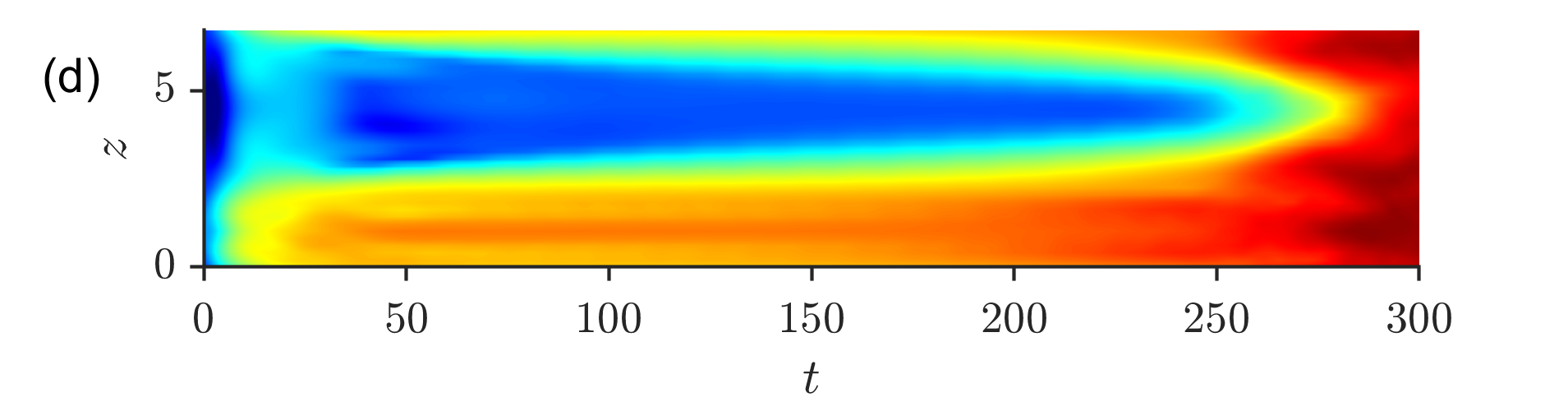} \\
	\includegraphics[width=0.45\textwidth]{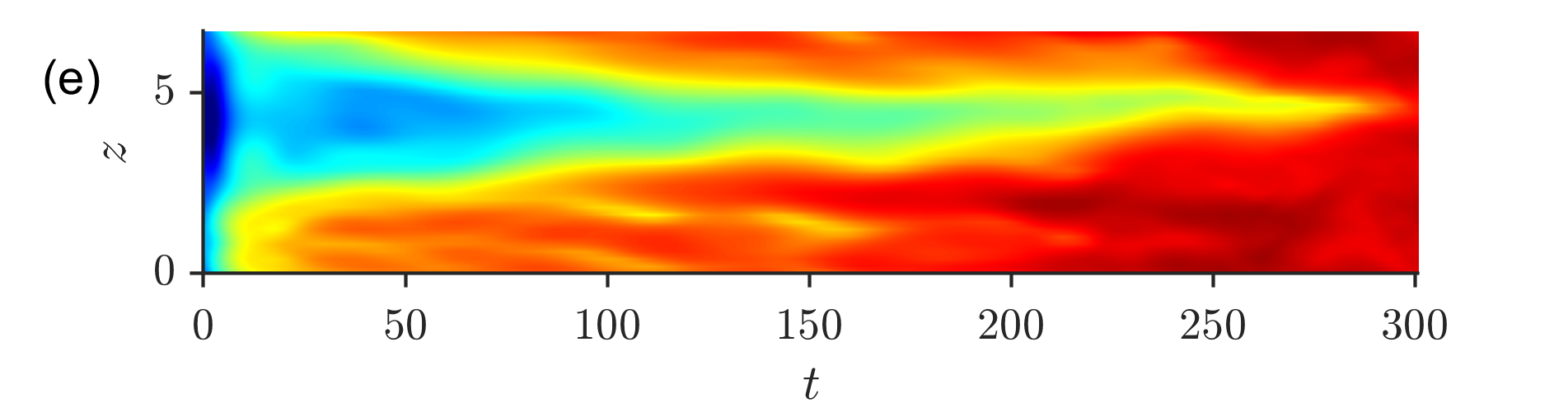}
	\includegraphics[width=0.45\textwidth]{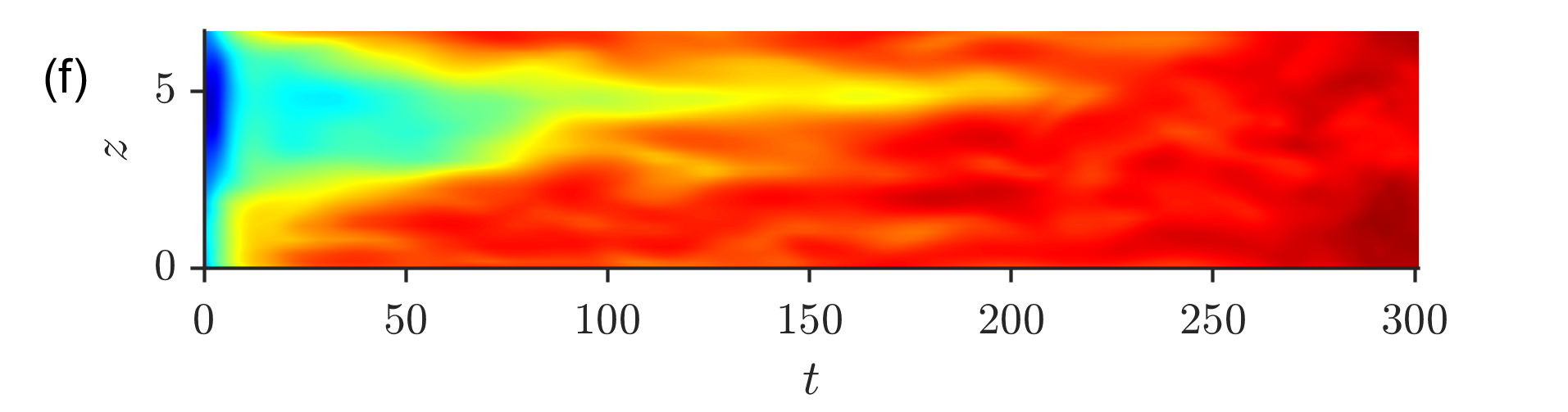} \\
	\includegraphics[width=0.45\textwidth]{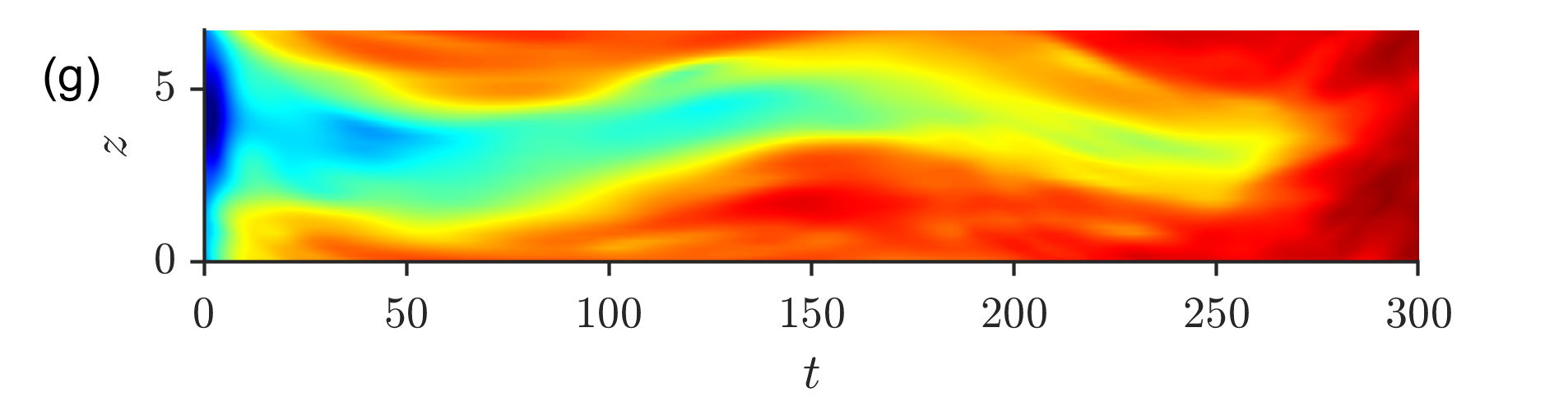}
	\includegraphics[width=0.45\textwidth]{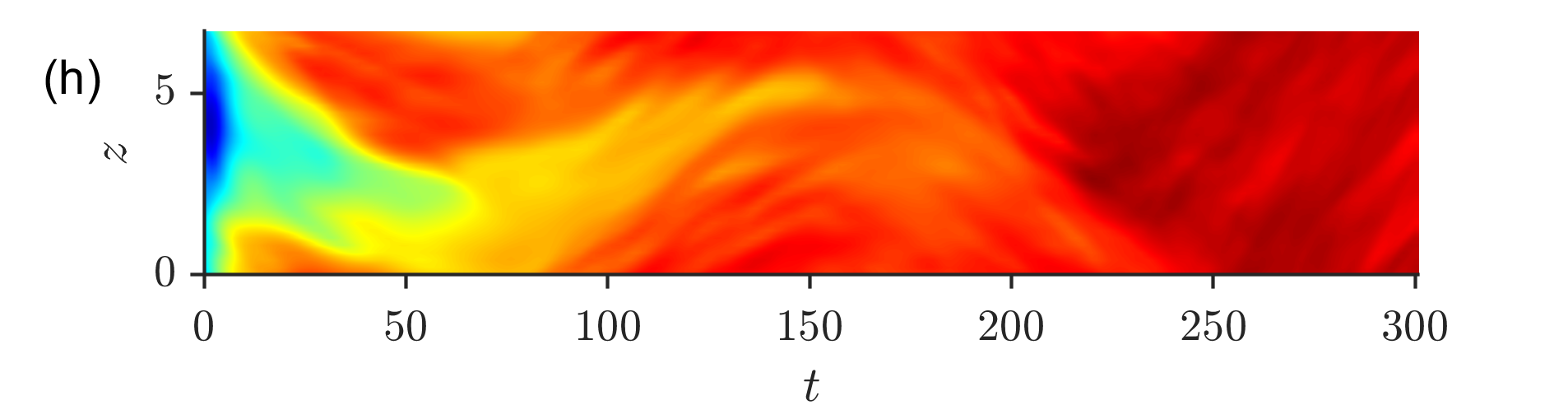}
	\end{minipage}%
	\begin{minipage}{0.1\textwidth}
	\includegraphics[width=0.95\textwidth]{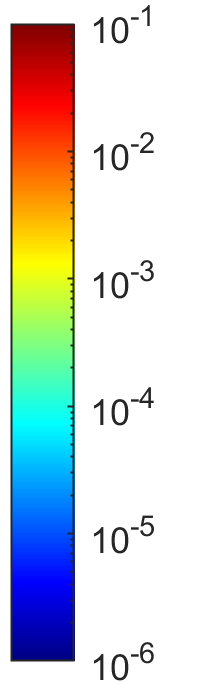}
	\end{minipage}
  \caption{(a) Upper bounds for the minimal seed turbulent kinetic energy $E_c$ against $\omega$ for $W_{osc} = 0$ (horizontal black line), 0.2 (blue) and 0.4 (red). Open red symbols are parameter values where the algorithm did not converge (see text). (b-h) Time-evolution of the $xy$-averaged turbulent kinetic energy of the minimal seed perturbation and its trajectory. The colour scheme is applied following a logarithmic scale over 5 orders of magnitude from $10^{-6}$ (dark blue) to 0.1 (dark red). The parameter values are (b) $W_{osc}=0$, (c,e,g) $W_{osc}=0.2$ and $\omega = 1/2$, $1/8$ and $1/32$ respectively, and (d,f,h) $W_{osc}=0.4$ and $\omega =1/2$, $1/8$ and $1/32$ respectively.}
\label{fig:minseeds}
\end{figure}

Minimal seeds were computed for $W_{osc}=0, 0.2$ and $0.4$ with $\omega = 1/2$, $1/8$, and $1/32$.
Figure \ref{fig:minseeds}(a) shows the upper bounds for $E_c$ found by iteratively solving the minimal seed nonlinear energy optimisation problem \citep[for a description of the optimisation problem and its solution, see][]{Kerswell2018}.
The turbulent kinetic energy for $W_{osc}=0$ and those indicated by filled symbols for $W_{osc} = 0.2$ and $0.4$ are all converged results.
Two of the results for $W_{osc}=0.4$ did not converge: with $\omega=1/8$ after 1538 iterations and with $\omega=1/32$ after 824 iterations, each iteration taking between 1 and 2 hours with 16 CPUs. In contrast, the minimal seed for $W_{osc}=0$ converged after 267 iterations.
Nonetheless, the turbulent kinetic energies reported in figure \ref{fig:minseeds} for the two non-converged cases are the smallest values for which an initial condition on the turbulent side of the edge manifold have been found; as such, they are still upper bounds on $E_c$, though unlike the converged cases we have no measure of how close the non-converged values are to $E_c$.

The results in figure \ref{fig:minseeds} indicate that the oscillation amplitude $W_{osc}$ has little effect on the minimal seed energy for $\omega=1/2$, echoing the results for the edge state energies shown in figure \ref{fig:edge_energy}(b). 
For $\omega = 1/8$ and $1/32$, a difference between $W_{osc}=0$ and $W_{osc}\neq 0$ is seen.
Unlike the edge state energies, it is evident that, as $\omega$ decreases, the minimal seed energy actually increases, representing a slight stabilisation of the laminar flow (albeit in a local optimal sense).
The minimal seed energy also increases as the oscillation amplitude $W_{osc}$ increases.
This is also different to edge state energy observations, and indicates that, as the flow becomes more nonlinearly stable (i.e. as $E_c$ increases), the energy gap between the minimal seed and the edge state decreases. 

The other panels in figure \ref{fig:minseeds} show the time-evolution of the $xy$-averaged turbulent kinetic energy of each minimal seed as it evolves along the edge manifold and eventually into the turbulent attractor.
Since the turbulent kinetic energy spans several orders of magnitude during each of these simulations, the colour scheme is applied in a logarithmic scale.
The dynamics for $W_{osc}=0$ (b) and for $\omega= 1/2$ with $W_{osc} = 0.2$ (c) and $0.4$ (d) are similar, consisting of a spanwise- (and streamwise-) localised initial condition which unwraps into a streamwise-aligned, spanwise-localised streaky vortex structure.
The spanwise extent of this structure slowly expands for around 200 time units, as the trajectory follows the edge of chaos, before rapidly breaking down and spreading in the spanwise direction as the trajectory moves along an unstable manifold of the edge and towards turbulence. This sequence of events qualitatively replicates those found for $W_{osc}=0$ and $\Rey=1000$ by \citet{Eaves2015} in their `wide' geometry, who also discuss the flow structures and their evolution in greater detail.

The \emph{initial} evolution ($t\lesssim 20$) of the minimal seed for $\omega=1/8$ in (e) and (f) is similar to that for $\omega = 1/2$ and $W_{osc} = 0$.
However, the spanwise-localised streaky vortex structure that the initial condition unwraps into is not streamwise-aligned, and the structure slowly drifts in the $z$-direction.
Every $2\pi/\omega\approx 50$ time units, the spanwise-extent of the perturbation increases a little, as can be seen at around $t=80$ and $130$.
Eventually, the perturbation encompasses the entire domain and fully developed turbulence is reached.
Finally, the dynamics for $\omega=1/32$ in (g) and (h) are similar to $W_{osc}=0$ in the \emph{initial} stages, but then undergo a large-amplitude meander in the spanwise direction with a period of approximately $2\pi/\omega \approx 200$ before breaking down to fully-developed turbulence.

\section{Computing the laminarization probability}

Both the edge state and the minimal seed are important to understand the dynamical mechanisms at play during transition to turbulence. However, they are local characteristics and do not allow us to appreciate the global structure of the edge.
To obtain a global characterization of the edge, we introduce the laminarization probability, which is the probability $P_{lam}(E)$ that a random perturbation (hereafter RP) from the laminar flow decays as a function of its initial energy $E$.
This quantity was first used in the uncontrolled plane Couette flow case as well as in the presence of spanwise wall oscillation for a very limited set of control parameter values \citep{Pershin2020}.
However, the algorithm for the approximation of $P_{lam}(E)$, described in that work and involving time-integration of 200 RPs per energy level $E$, is insufficiently fast to enable efficient optimization of the control parameters.
We propose here a solution based on the Bayesian estimation of the laminarization probability that allows for the computation of a significantly reduced number of simulations, and hence efficient optimisation.

\subsection{Computing the laminarization probability using Bayesian inference}
\label{subsec:bayesian_inference}

First, we derive the distribution of the laminarization probability from observations carried out during the numerical simulation of a finite number of RPs.
This distribution allows for the estimation of the laminarization probability as a function of the number of laminarizing RPs and the size of the observation sample.
Let us consider the laminarization probability $P_{lam} = P_{lam}(E)$, a function of the energy level $E$, as a random variable.
We also introduce $\boldsymbol{R} = (R_1, R_2, \dots, R_N)$, a sample consisting of $N$ elements, with each $R_i$ associated with an RP of the same energy level and taking Boolean values: $R_i = 1$ if laminarization is observed or $R_i = 0$ if transition is observed. Since turbulence is always transient for the given flow configuration, we assume that the flow transitions to turbulence if it does not laminarize for at least $400$ time units (see \cite{Pershin2020} for further details).
The probability mass function associated with event $R_i = r$, knowing that $P_{lam} = p$, is:
\begin{equation}
\mathbb{P}(R_i = r | P_{lam} = p) = \left\{
    \begin{array}{ll}
    p, & r = 1 \\
    1 - p, & r = 0
    \end{array} \right. .
\end{equation}

Suppose that the sample $\boldsymbol{R}$ takes particular values $\boldsymbol{r} = (r_1, r_2, \dots, r_N)$ and presents $l = \sum_{i = 1}^{N} r_i$ laminarization events.
Since all the simulations are independent, the probability of observing such sample values, given the laminarization probability $P_{lam} = p$, is:
\begin{equation}
\label{eq:likelihood}
\mathbb{P}(\boldsymbol{R} = \boldsymbol{r} | P_{lam} = p) = p^{l} (1 - p)^{N - l}.
\end{equation}
We can now obtain the probability density function $f_{P_{lam}}$ for $P_{lam}$ for a given sample $\boldsymbol{R} = \boldsymbol{r}$ by applying Bayes' theorem:
\begin{equation}
\label{eq:posterior_general}
f_{P_{lam}}(p | \boldsymbol{R} = \boldsymbol{r}) = \frac{\mathbb{P}(\boldsymbol{R} = \boldsymbol{r} | P_{lam} = p) \cdot f_{P_{lam}}(p)}{\mathbb{P}(\boldsymbol{R} = \boldsymbol{r})}, \quad p \in (0; 1),
\end{equation}
where $f_{P_{lam}}(p | \boldsymbol{R} = \boldsymbol{r})$ is referred to as the \textit{posterior distribution} for $p$, $f_{P_{lam}}(p)$ is its \textit{prior distribution}, $\mathbb{P}(\boldsymbol{R} = \boldsymbol{r} | P_{lam} = p)$ is called the \textit{likelihood} and $\mathbb{P}(\boldsymbol{R} = \boldsymbol{r})$ is the \textit{marginal likelihood}, which can be expressed in terms of the likelihood function and the prior distribution:
\begin{equation}
  \mathbb{P}(\boldsymbol{R} = \boldsymbol{r}) = \int_0^1 \mathbb{P}(\boldsymbol{R} = \boldsymbol{r} | P_{lam} = p) f_{P_{lam}}(p) \mathrm{d}p.
\end{equation}

Choosing the prior distribution is a crucial step in Bayesian inference and must encompass all our knowledge about $P_{lam}$.
When no such knowledge is available, the prior distribution is usually taken to be the least informative among the whole space of admissible distributions of $P_{lam}$.
Here, more than one possibility is available.
Two of the most widely used options are the Jeffreys' prior distribution,
\begin{equation}
\label{eq:jeffreys_prior}
f_{P_{lam}}(p) = \frac{1}{\pi \sqrt{p(1-p)}}, \quad p \in (0; 1),
\end{equation}
and the uniform prior distribution, $f_{P_{lam}}(p) = 1$, for $p \in (0;1)$ (see Appendix \ref{app:priors} for details and justifications of such choices).
Whilst choosing either of two prior distributions does not significantly affect the posterior distribution for large $N$, it does have significant influence for $N \le O(10)$.
Preliminary analysis of the effect of the choice of a prior distribution for $W_{osc} = 0$ and $(W_{osc}, \omega) = (0.3,1/16)$ shows that the use of the uniform prior distribution makes the probabilistic model conservative around extreme values of $P_{lam}$ so that it tends to predict values for $P_{lam}$ closer to $0.5$ than they actually are.

\begin{figure}
    \includegraphics[width=1.0\textwidth]{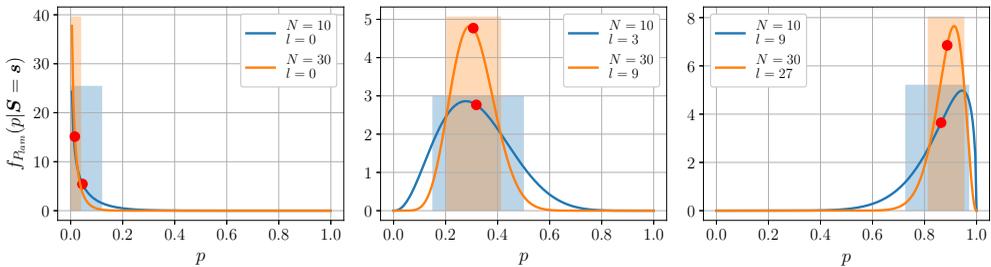}
    \caption{Examples of posterior distributions of the form of equation (\ref{eq:posterior}) plotted for $N = 10$ (blue) and $N = 30$ (orange) and three values of the fraction of laminarizing RPs: $l/N = 0$ (left), $l/N = 3/10$ (middle) and $l/N = 9/10$ (right). Red dots and shaded regions denote the means and interdecile ranges of the corresponding distributions respectively.}
    \label{fig:posterior_examples}
\end{figure}

The Jeffreys' prior distribution, in contrast, does not seem to introduce any such significant bias;  this is confirmed during the validation of the described approach by generating a large number of random samples of RPs from an available dataset and then predicting the laminarization probability (see subsection \ref{subsec:validation} for details).
As a result, we opted for the Jeffreys' prior distribution.
Substituting equation (\ref{eq:jeffreys_prior}) into equation (\ref{eq:posterior_general}) yields the posterior distribution in the form of the beta distribution:
\begin{equation}
\label{eq:posterior}
f_{P_{lam}}(p | \boldsymbol{R} = \boldsymbol{r}) = \frac{p^{l - 1/2} (1 - p)^{N - l - 1/2}}{\mbox{B}(l + 1/2, N - l + 1/2)}, \quad p \in (0; 1),
\end{equation}
where $\mbox{B}(\cdot, \cdot)$ is the beta function.
Now we can readily get a point estimate of the laminarization probability $\bar{P}_{lam}$ as the expectation of $P_{lam}$ calculated with respect to the posterior distribution (\ref{eq:posterior}):
\begin{equation}
\label{eq:p_lam_expect}
\bar{P}_{lam} = \int_0^1 p f_{P_{lam}}(p | \boldsymbol{R} = \boldsymbol{r}) \mathrm{d}p = \frac{l + 1/2}{N + 1}.
\end{equation}
As a convenient consequence of the use of Bayesian inference, the interval estimates (e.g., interquartile and interdecile ranges) assessing a range of values within which $P_{lam}$ is likely to be located, can also be found directly from the posterior distribution.
We show examples of posterior distributions with their point and interval characteristics in figure \ref{fig:posterior_examples} for different combinations of the number of laminarizing events $l$ and the sample size $N$.
Posterior distributions become more peaked as the number of RPs per energy level $N$ is increased, thereby reflecting the fact that the probability of mispredicting $P_{lam}$ approaches zero as $N \to \infty$.

\subsection{Numerical methodology}

\begin{figure}
    \centering
    \begin{tikzpicture}[x=0.9cm, y=0.2cm]
        \def\Xmin{0}
        \def\Xmax{14}
        \def\Ymin{0}
        \def\Ymax{50}

        \def\Emin{1}
        \def\Emax{13}
        \def\Eoffset{0.8}
        \def\Edeltabetween{1.6}
        \def\Edeltain{0.5}

        \def\RPmin{34}
        \def\RPmax{48}
        \def\RPdelta{1.5}

        \def\Yposteriors{\RPmin - 12}
        \def\Yfitting{\Ymin + 3}


        \draw[ultra thick] (\Emin, \RPmin) -- (\Emax, \RPmin);

        \draw [thick, fill=cyan] (\Emin + \Eoffset, \RPmin) rectangle (\Emin + \Eoffset + \Edeltain, \RPmin + \RPdelta);
        \draw [thick, fill=cyan] (\Emin + \Eoffset, \RPmin + \RPdelta) rectangle (\Emin + \Eoffset + \Edeltain, \RPmin + 2*\RPdelta);
        \draw [thick, fill=cyan] (\Emin + \Eoffset, \RPmin + 2*\RPdelta) rectangle (\Emin + \Eoffset + \Edeltain, \RPmin + 3*\RPdelta);
        \draw [thick, fill=cyan] (\Emin + \Eoffset, \RPmin + 3*\RPdelta) rectangle (\Emin + \Eoffset + \Edeltain, \RPmin + 4*\RPdelta);
        \draw [thick, fill=cyan] (\Emin + \Eoffset, \RPmin + 4*\RPdelta) rectangle (\Emin + \Eoffset + \Edeltain, \RPmin + 5*\RPdelta);
        \draw [thick, fill=cyan] (\Emin + \Eoffset, \RPmin + 5*\RPdelta) rectangle (\Emin + \Eoffset + \Edeltain, \RPmin + 6*\RPdelta);
        \draw [thick, fill=cyan] (\Emin + \Eoffset, \RPmin + 6*\RPdelta) rectangle (\Emin + \Eoffset + \Edeltain, \RPmin + 7*\RPdelta);
        \draw [thick, fill=cyan] (\Emin + \Eoffset, \RPmin + 7*\RPdelta) rectangle (\Emin + \Eoffset + \Edeltain, \RPmin + 8*\RPdelta);
        \draw [thick, fill=cyan] (\Emin + \Eoffset, \RPmin + 8*\RPdelta) rectangle (\Emin + \Eoffset + \Edeltain, \RPmin + 9*\RPdelta);
        \draw [thick, fill=cyan] (\Emin + \Eoffset, \RPmin + 9*\RPdelta) rectangle (\Emin + \Eoffset + \Edeltain, \RPmin + 10*\RPdelta);

        \draw [thick, fill=cyan] (\Emin + \Eoffset + \Edeltain + \Edeltabetween, \RPmin) rectangle (\Emin + \Eoffset + 2*\Edeltain + \Edeltabetween, \RPmin + \RPdelta);
        \draw [thick, fill=purple] (\Emin + \Eoffset + \Edeltain + \Edeltabetween, \RPmin + \RPdelta) rectangle (\Emin + \Eoffset + 2*\Edeltain + \Edeltabetween, \RPmin + 2*\RPdelta);
        \draw [thick, fill=cyan] (\Emin + \Eoffset + \Edeltain + \Edeltabetween, \RPmin + 2*\RPdelta) rectangle (\Emin + \Eoffset + 2*\Edeltain + \Edeltabetween, \RPmin + 3*\RPdelta);
        \draw [thick, fill=cyan] (\Emin + \Eoffset + \Edeltain + \Edeltabetween, \RPmin + 3*\RPdelta) rectangle (\Emin + \Eoffset + 2*\Edeltain + \Edeltabetween, \RPmin + 4*\RPdelta);
        \draw [thick, fill=cyan] (\Emin + \Eoffset + \Edeltain + \Edeltabetween, \RPmin + 4*\RPdelta) rectangle (\Emin + \Eoffset + 2*\Edeltain + \Edeltabetween, \RPmin + 5*\RPdelta);
        \draw [thick, fill=cyan] (\Emin + \Eoffset + \Edeltain + \Edeltabetween, \RPmin + 5*\RPdelta) rectangle (\Emin + \Eoffset + 2*\Edeltain + \Edeltabetween, \RPmin + 6*\RPdelta);
        \draw [thick, fill=purple] (\Emin + \Eoffset + \Edeltain + \Edeltabetween, \RPmin + 6*\RPdelta) rectangle (\Emin + \Eoffset + 2*\Edeltain + \Edeltabetween, \RPmin + 7*\RPdelta);
        \draw [thick, fill=cyan] (\Emin + \Eoffset + \Edeltain + \Edeltabetween, \RPmin + 7*\RPdelta) rectangle (\Emin + \Eoffset + 2*\Edeltain + \Edeltabetween, \RPmin + 8*\RPdelta);
        \draw [thick, fill=cyan] (\Emin + \Eoffset + \Edeltain + \Edeltabetween, \RPmin + 8*\RPdelta) rectangle (\Emin + \Eoffset + 2*\Edeltain + \Edeltabetween, \RPmin + 9*\RPdelta);
        \draw [thick, fill=cyan] (\Emin + \Eoffset + \Edeltain + \Edeltabetween, \RPmin + 9*\RPdelta) rectangle (\Emin + \Eoffset + 2*\Edeltain + \Edeltabetween, \RPmin + 10*\RPdelta);

        \draw [thick, fill=purple] (\Emax - \Eoffset - \Edeltain, \RPmin) rectangle (\Emax - \Eoffset, \RPmin + \RPdelta);
        \draw [thick, fill=cyan] (\Emax - \Eoffset - \Edeltain, \RPmin + \RPdelta) rectangle (\Emax - \Eoffset, \RPmin + 2*\RPdelta);
        \draw [thick, fill=purple] (\Emax - \Eoffset - \Edeltain, \RPmin + 2*\RPdelta) rectangle (\Emax - \Eoffset, \RPmin + 3*\RPdelta);
        \draw [thick, fill=cyan] (\Emax - \Eoffset - \Edeltain, \RPmin + 3*\RPdelta) rectangle (\Emax - \Eoffset, \RPmin + 4*\RPdelta);
        \draw [thick, fill=purple] (\Emax - \Eoffset - \Edeltain, \RPmin + 4*\RPdelta) rectangle (\Emax - \Eoffset, \RPmin + 5*\RPdelta);
        \draw [thick, fill=purple] (\Emax - \Eoffset - \Edeltain, \RPmin + 5*\RPdelta) rectangle (\Emax - \Eoffset, \RPmin + 6*\RPdelta);
        \draw [thick, fill=purple] (\Emax - \Eoffset - \Edeltain, \RPmin + 6*\RPdelta) rectangle (\Emax - \Eoffset, \RPmin + 7*\RPdelta);
        \draw [thick, fill=cyan] (\Emax - \Eoffset - \Edeltain, \RPmin + 7*\RPdelta) rectangle (\Emax - \Eoffset, \RPmin + 8*\RPdelta);
        \draw [thick, fill=purple] (\Emax - \Eoffset - \Edeltain, \RPmin + 8*\RPdelta) rectangle (\Emax - \Eoffset, \RPmin + 9*\RPdelta);
        \draw [thick, fill=purple] (\Emax - \Eoffset - \Edeltain, \RPmin + 9*\RPdelta) rectangle (\Emax - \Eoffset, \RPmin + 10*\RPdelta);

        \draw [decorate,decoration={brace,amplitude=10pt},xshift=5pt,yshift=0pt] (\Emax - \Eoffset, \RPmin + 10*\RPdelta - 0.4) -- (\Emax - \Eoffset, \RPmin + 0.4) node [black, midway, xshift=27pt] {$N$ RPs};

        \node at (\Emin + 0.5*\Eoffset, \RPmin + 5*\RPdelta) [rotate=90] {$\boldsymbol{R} = (1, 1, 1, \dots, 1)$};
        \node at (\Emin + 0.5*\Eoffset + \Edeltain + \Edeltabetween, \RPmin + 5*\RPdelta) [rotate=90] {$\boldsymbol{R} = (1, 0, 1, \dots, 1)$};
        \node at (\Emax - 1.5*\Eoffset - \Edeltain, \RPmin + 5*\RPdelta) [rotate=90] {$\boldsymbol{R} = (0, 1, 0, \dots, 0)$};

        \node at (\Emin + \Eoffset + 0.5*\Edeltain, \RPmin - 2) {$E^{(1)}$};
        \node at (\Emin + \Eoffset + 1.5*\Edeltain + \Edeltabetween, \RPmin - 2) {$E^{(2)}$};
        \node at (\Emax - \Eoffset - 0.5*\Edeltain, \RPmin - 2) {$E^{(j)}$};

        \draw [pil, ->] (\Emin + \Eoffset + 0.5*\Edeltain, \RPmin - 3.5) -- (\Emin + \Eoffset + 0.5*\Edeltain, \Yposteriors + 4);
        \draw [pil, ->] (\Emin + \Eoffset + 1.5*\Edeltain + \Edeltabetween, \RPmin - 3.5) -- (\Emin + \Eoffset + 1.5*\Edeltain + \Edeltabetween, \Yposteriors + 4);
        \draw [pil, ->] (\Emax - \Eoffset - 0.5*\Edeltain, \RPmin - 3.5) -- (\Emax - \Eoffset - 0.5*\Edeltain, \Yposteriors + 4);

        \node[rectangle, draw] (POSTERIOR_ONE) at (\Emin + \Eoffset + 0.5*\Edeltain, \Yposteriors) {\includegraphics[width=0.1\textwidth]{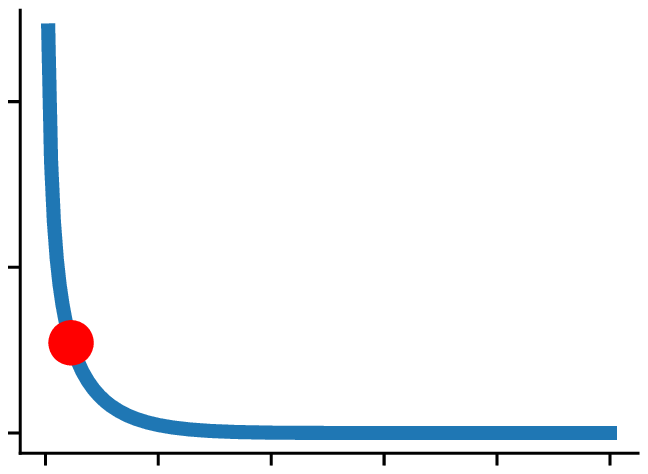}};
        \node[rectangle, draw] (POSTERIOR_TWO) at (\Emin + \Eoffset + 1.5*\Edeltain + \Edeltabetween, \Yposteriors) {\includegraphics[width=0.1\textwidth]{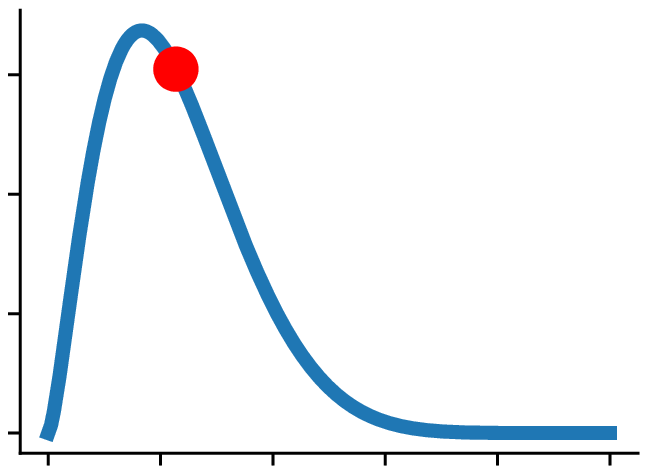}};
        \node[rectangle, draw] (POSTERIOR_LAST) at (\Emax - \Eoffset - 0.5*\Edeltain, \Yposteriors) {\includegraphics[width=0.1\textwidth]{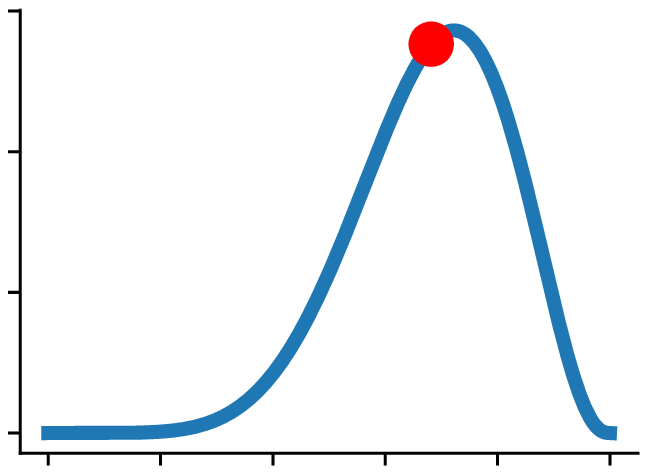}};

        \node (FITTING) at ({0.5*(\Emin + \Emax)}, \Yfitting) {\includegraphics[width=0.9\textwidth]{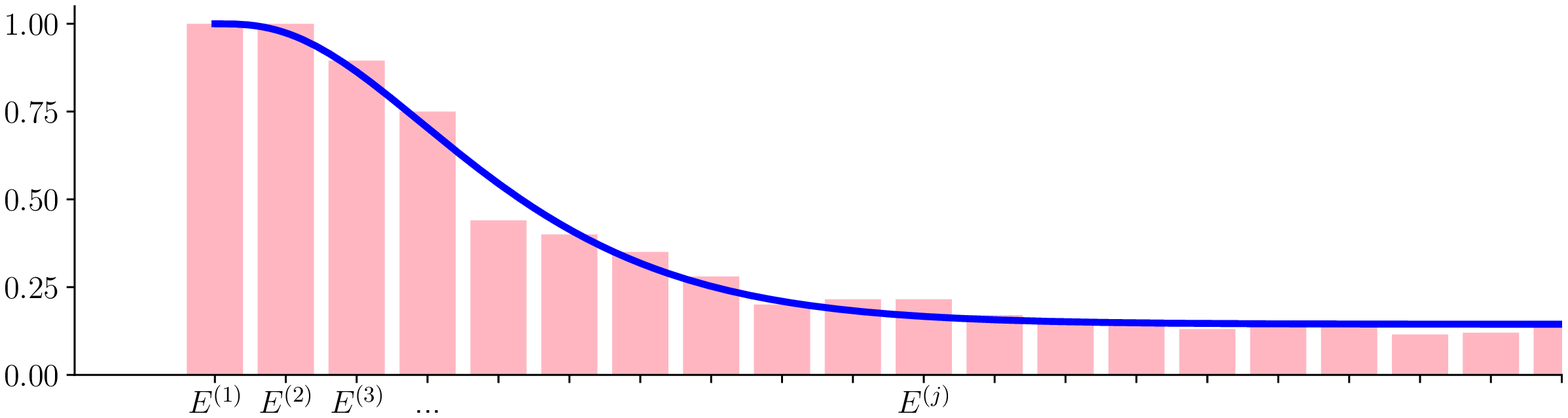}};

        \draw [pil, ->] (\Emin + \Eoffset + 0.5*\Edeltain, \Yposteriors - 4) -- (\Emin + \Eoffset + 0.5*\Edeltain + 0.25, \Yposteriors - 11.4) node [midway, left] (EXP_ONE) {$\bar{P}_{lam}(E^{(1)})$};
        \draw [pil, ->] (\Emin + \Eoffset + 1.5*\Edeltain + \Edeltabetween, \Yposteriors - 4) -- (\Emin + \Eoffset + 0.5*\Edeltain + 0.9, \Yposteriors - 11.4) node [midway, right] (EXP_TWO) {$\bar{P}_{lam}(E^{(2)})$};
        \draw [pil, ->] (\Emax - \Eoffset - 0.5*\Edeltain, \Yposteriors - 4) -- ({0.5*(\Emin + \Emax) + 1.2}, \Ymin + 0.4) node [midway, right] (EXP_LAST) {$\bar{P}_{lam}(E^{(j)})$};

        \node[rounded corners, fill=lightgray, text width=4cm] at ({0.5*(\Emin+\Emax) + 1}, {0.5*(\RPmin+\RPmax)}) {\scriptsize \textbf{Step 1:} sampling $\boldsymbol{R}$ for each energy level};
        \node[rounded corners, fill=lightgray, text width=4cm] at ({0.5*(\Emin+\Emax) + 1}, \Yposteriors) {\scriptsize \textbf{Step 2:} computing posterior distributions $f_{P_{lam}}(p | \boldsymbol{R} = \boldsymbol{r})$};
        \node[rounded corners, fill=lightgray, text width=3.5cm] at ({0.5*(\Emin+\Emax) - 0.2}, \Ymin + 8.5) {\scriptsize \textbf{Step 3:} computing point estimates of $P_{lam}$ with respect to the posteriors};
    \end{tikzpicture}
    \caption{Schematic procedure used to approximate the laminarization probability. In step 1, we build a sample $\boldsymbol{R}$ for each RP energy level and determine the number of laminarizing events. Blue (resp. red) blocks correspond to laminarizing (resp. transitioning) RPs. Next, in step 2, we use the count obtained in each of these samples to compute the associated posterior distribution $f_{P_{lam}}(p | \boldsymbol{R} = \boldsymbol{r})$ based on expression (\ref{eq:posterior}) for each energy level. We then compute the point estimates of the laminarization probability based on expression (\ref{eq:p_lam_expect}) for each energy level. These points are shown using red dots on the step 2 curves and are shown on the bottom panel as pink bars. The approximate dependence of the laminarization probability on the RP energy is then obtained in step 3 via least-squares fitting based on the fitting function $p(E)$ (blue curve in the bottom plot) introduced in (\ref{eq:fittingfunc}). More details on the final fitting are shown in figure \ref{fig:fitting_sketch}.}
    \label{fig:estimation_schema}
\end{figure}

Our methodology is summarized in figure \ref{fig:estimation_schema}.
\begin{figure}
    \includegraphics[width=1.0\textwidth]{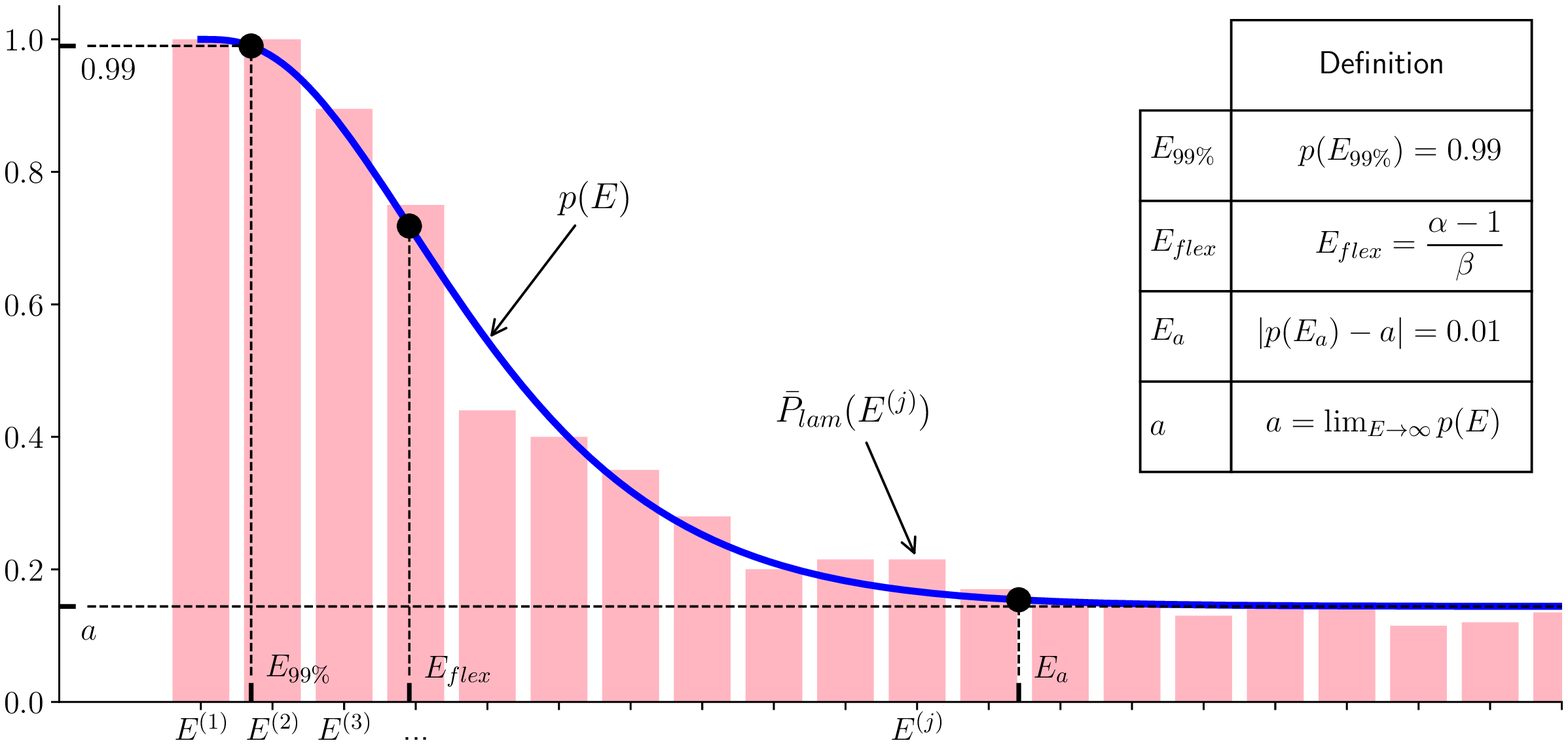}
    \caption{Examples of estimated laminarization probabilities $P_{lam}$ (pink bars) plotted together with the associated fitting function $p(E) = 1 - (1 - a)\gamma(\alpha, \beta E)$, where $\gamma(\alpha, \beta E)$ is the lower incomplete gamma function (blue curve), together with its characterizing quantities (indicated with black dots on the plot) defined in the table.}
    \label{fig:fitting_sketch}
\end{figure}
Since the laminarization probability depends on the RP kinetic energy, we first discretize the relevant energy range by considering $40$ energy levels $E^{(j)}, j = 1, \dots, 40$, equispaced between $0$ and $E_{max} = 0.04$ (see \citet{Pershin2020}).
For each energy level $E^{(j)}$, we then perform $N$ numerical simulations using $N$ generated RPs as initial conditions (see Appendix \ref{app:rp_generation} for the details of the generation procedure) and count the number of laminarization events $l$ (step 1 in figure \ref{fig:estimation_schema}).
Next, we compute the posterior distribution (\ref{eq:posterior}) based on the values of $l$ and $N$ (step 2 in figure \ref{fig:estimation_schema}).
Once posterior distributions are built for all energy levels, the point estimate of the laminarization probability are calculated using equation (\ref{eq:p_lam_expect}) at each energy level (step 3 in figure \ref{fig:estimation_schema}).
Finally, an approximation of the dependence of the laminarization probability on the RP kinetic energy is obtained by least-squares fitting of the function 
\begin{equation}
\label{eq:fittingfunc}
    p(E) = 1 - (1 - a)\gamma(\alpha, \beta E)
\end{equation}
to the point estimates of $\bar{P}_{lam}(E^{(j)})$, where $\gamma(\alpha, \beta E)$ is the lower incomplete gamma function.
This function possesses characteristics that make it a good fitting choice \citep{Pershin2020}.

We find it convenient to characterize the fitting curve of the laminarization probability by four scalar quantities: the value of the RP energy at which the laminarization probability reaches $99\%$, $E_{99\%}$; the value of the RP energy at which the fitting function undergoes its inflection point, $E_{flex}$; the asymptotic value $a$ of the laminarization probability; and the value of the RP energy at which the laminarization probability is $1\%$ away from $a$, $E(a)$.
These definitions are illustrated in figure \ref{fig:fitting_sketch}.
The quantity $E_{99\%}$ can be thought of as the perturbation energy beyond which there is a non-negligible probability of observing transition to turbulence.
From the perspective of the assessment of control strategies, this value might be considered as a practical substitute to the minimal seed energy as it is less dependent on perturbing the flow with the minimal seed's specific spatial structure, or of a random perturbation lying within the likely negligible state space volume occupied by transitional initial conditions nearby to the minimal seed.
As the perturbation kinetic energy is increased, the fitting function decreases monotonically and undergoes an inflection point at $E_{flex}$ to approach its asymptotic value $a$.
We can consider that the laminarization probability plateaus when the RP energy reaches $E_{a}$.
Even though this choice for the fitting function proves satisfactory for the flow configurations and perturbation energy ranges under consideration, it might not be appropriate elsewhere.
For example, the laminarization probability may return back to $1$ as $E \to \infty$ owing to the upper edge of chaos \citep{Budanur2020} so that modelling the laminarization probability over the entire energy range would require a different fitting function.


To be able to compare the efficiency of different control strategies, it is important to remember that the kinetic energy of the initial condition depends on the origin and type of the disturbance to which the flow is subjected.
We therefore introduce the probability density function $f_E(E)$ associated with the probability that a disturbance to the flow is created with energy $E$.
This allows the definition of the \textit{laminarization score}: 
\begin{equation}
\label{eq:lam_score}
S = \int_0^{E_{max}} p(E) f_E(E) dE,
\end{equation}
which represents the expected value of the laminarization probability $p(E)$ assuming that the perturbation energy is distributed according to $f_E(E)$.
In other words, the laminarization score is the probability of observing laminarization in a configuration where the laminarization probability is $p(E)$ and where the probability that a perturbation is generated at energy $E$ is $f_E(E)$.
We may infer this distribution from experimental observations or by using \emph{a priori} knowledge of the perturbation generation mechanisms.
In this work, we assume no prior knowledge of this kind and consider two potentially useful distributions up to a maximum value $E_{max}$: the uniform distribution $f_E^{(uni)}(E)$, which implies that the source of disturbances behaves like a white noise emitter, and the exponential distribution $f_E^{(exp)}(E)$, adjusted to finite support, to model cases where small-energy disturbances are more commonly generated than large-energy ones:
\begin{eqnarray}
f_E^{(uni)}(E) &=& \frac{1}{E_{max}}, \\
f_E^{(exp)}(E) &=& \frac{\lambda}{1 - e^{-\lambda E_{max}}} e^{-\lambda E},
\end{eqnarray}
where $\lambda \approx E_{avg}^{-1}$ approximates the inverse of the average energy for sufficiently large $E_{max} / E_{avg}$.

\subsection{Approximation errors and confidence intervals}
\label{subsec:validation}

To test our approach, we first estimate $P_{lam}(E^{(j)})$ for the uncontrolled case ($W_{osc} = 0$) and the  controlled case: $W_{osc} = 0.3, \omega=1/16$.
A large number of RPs ($200$ per energy level) were previously used in both cases to accurately approximate the laminarization probability \citep{Pershin2020}.
These results provide fitting functions against which our new Bayesian approach can be tested.
We will refer to the (converged) fitting function obtained in this previous work as $p_{acc}(E)$.
To test whether we can obtain acceptable estimates of the laminarization probability using a small-size sample, we randomly draw, with replacement, $1000$ samples of $N$ RPs per energy level.
For each sample, we perform the procedure summarized in figure \ref{fig:estimation_schema}.
From the resulting fitting function, $p(E)$, we extract parameters $a$, $\alpha$ and $\beta$.
This allows us to assess how the errors involved in approximating the accurate fitting function $p_{acc}(E)$ scale with $N$.
In particular, we track the relative errors in our estimates for  $E_{flex}$, $E_a$ and with $S$ subject to the uniform distribution $f_E^{(uni)}(E)$.
These errors are denoted $e_{flex}$, $e_{E_a}$ and $e_{S}$ respectively.

\begin{figure}
  \includegraphics[width=1.0\textwidth]{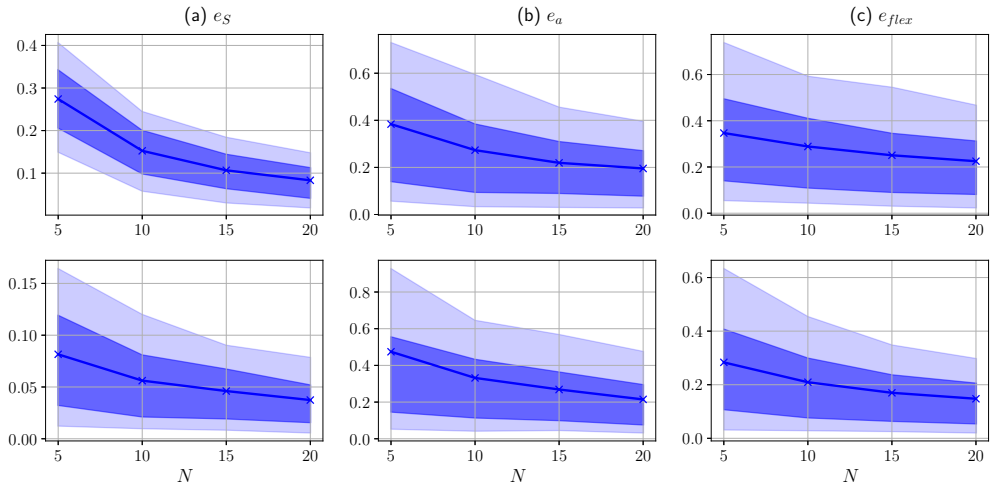}
  \caption{Relative errors in $S$, $E_{flex}$ and $E_a$, denoted $e_S$, $e_{flex}$ and $e_{a}$, as a function of $N$. These errors were computed by estimating the fitting function $p(E)$ using $N$ randomly selected RPs per energy level. Errors associated with the uncontrolled (controlled) system are shown in the top (bottom) row. The parameter values of the control are $W_{osc} = 0.3, \omega = 1/16$. The dark (resp. light) blue bands correspond to the interquartile (resp. interdecile) ranges. Dark blue lines show the expectations of associated errors as functions of $N$. 
  }
  \label{fig:p_est_errors}
\end{figure}

The dependence of these errors on $N$ for both the uncontrolled and the controlled cases is displayed in figure \ref{fig:p_est_errors}.
The largest relative errors are observed for $E_a$ and $E_{flex}$, exceeding $0.7$ in certain cases.
However, because $S$ is an integral quantity, inaccurate estimates of $E_a$ and $E_{flex}$ do not necessarily lead to a large error in $S$.
Indeed, one can observe that, for $N=10$, the relative error $e_{S}$ is lower than $0.25$ for $90\%$ of the samples in the uncontrolled cases, and is lower than $0.12$ for $90\%$ of the samples in the controlled case.
Such a difference between the uncontrolled and controlled cases is explained by the fact that with small sample sizes, relative error degrades compared with absolute error as the accurate value of $P_{lam}$ approaches zero, which is indeed the case for the majority of the energy levels for $W_{osc}=0$.

\begin{figure}
  \includegraphics[width=1.0\textwidth]{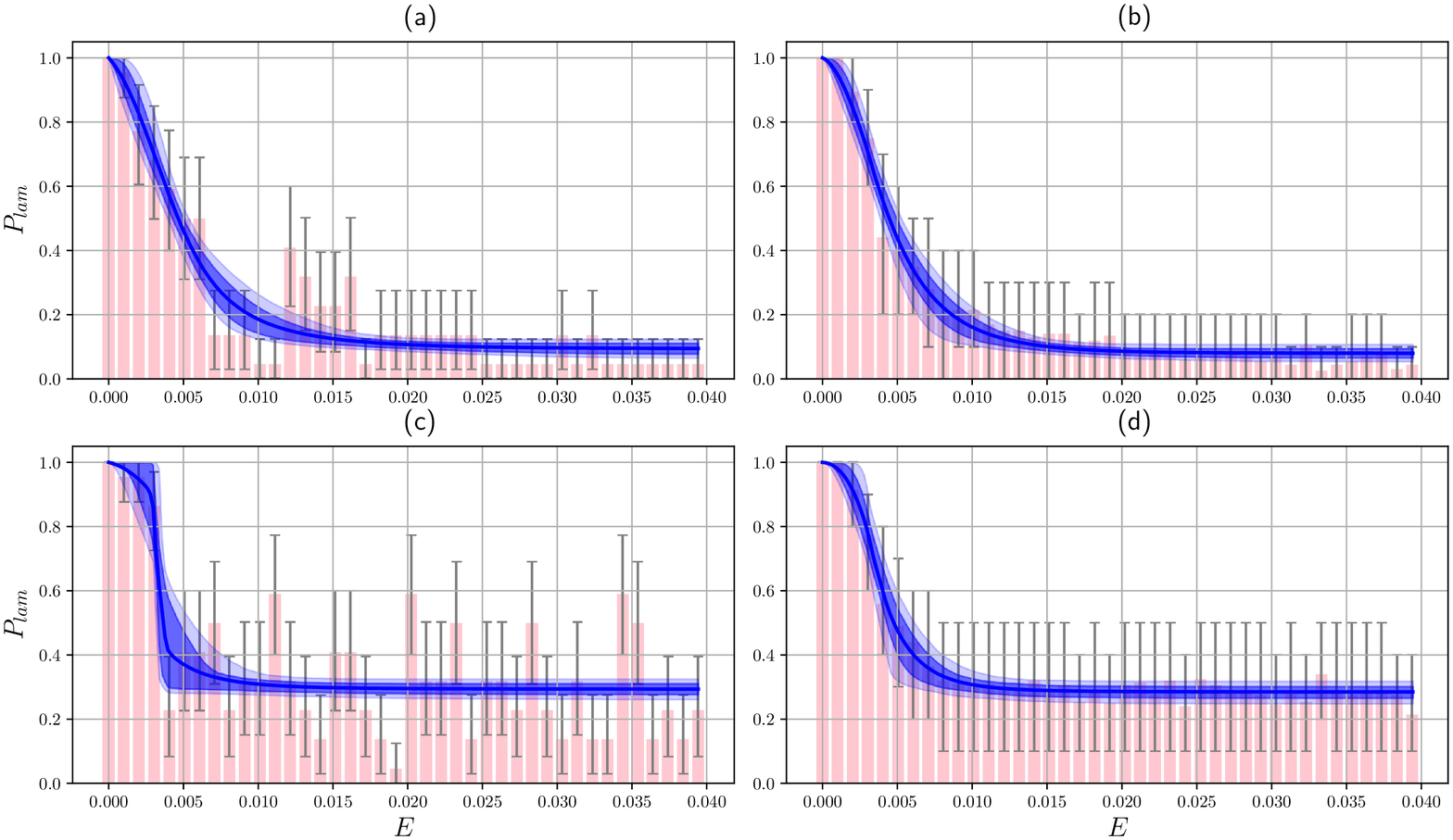}
  \caption{Expectations (solid lines) for the distribution of fitting functions and their interquartile (dark bands) and interdecile (light bands) ranges for the uncontrolled (a, b) and controlled (c, d) test cases. Left plots were computed using a single sample of $N = 10$ RPs per energy level, Bayesian estimation of the laminarization probability (pink bars) and drawing a large number of samples from the posterior distributions to build the associated interquartile and interdecile ranges (see the text for details). Right plots were computed by drawing $1000$ samples, where each sample corresponds to $10$ RPs per energy level, from a large database of RPs with replacement and then building a distribution of fitting functions each of which was obtained following the procedure illustrated in figure \ref{fig:estimation_schema}. Pink bars in the right plots correspond to the laminarization probability $P_{lam}$ estimated for all available RPs. Gray error bars correspond to the interdecile range of the laminarization probability estimated using Bayesian inference (left) or subsampling (right). 
  }
  \label{fig:p_lam_subsampling}
\end{figure}

We also kept track of the expectation and the interquartile and interdecile ranges of the distribution of fitting function $p(E)$ for $N=10$ and report the results in figure \ref{fig:p_lam_subsampling}(b, d).
The interdecile range will hereafter be treated as the confidence range or, equivalently, confidence band for the fitting function.
The interdecile range for $p(E)$ has typical amplitude around $0.02$, so the distribution for the fitting function is fairly concentrated around its expectation.
Some exceptions are observed for very small ($E \lesssim 3 \times 10^{-3}$) and moderate ($5 \times 10^{-3} \lesssim E \lesssim 9 \times 10^{-3}$) energy levels where this variation goes beyond $0.1$.
These are explained by the large variations in the coefficients $\alpha$ and $\beta$ obtained from different samples and are reflected in the relatively large values of errors $e_{a}$ and $e_{flex}$ in figure \ref{fig:p_est_errors}.
Given these observations, figure \ref{fig:p_lam_subsampling}(b, d) suggests that the laminarization probability can be reliably approximated using as few as $N=10$ RPs per energy level.
With such a sample size, the laminarization score $S$, a key indicator for the control efficiency, is expected $90\%$ of the time to take values yielding a relative error lower than $0.25$ .
This allows for an order-of-magnitude reduction of the number of simulations required to approximate the laminarization probability, which in turn allows the investigation of wider ranges of control parameter values without increasing cost.

As only $N = 10$ RPs per energy level are sufficient for a good approximation of the laminarization probability and the laminarization score, we can employ the Bayesian approach described in section \ref{subsec:bayesian_inference} for this estimation.
As has already been mentioned, by providing posterior distributions for $P_{lam}$, this approach not only yields the estimated point values of the laminarization probability $\bar{P}_{lam}$, but also makes uncertainty quantification possible by computing the associated confidence intervals given only a single sample. The provision of these uncertainty bands makes the Bayesian approach significantly more powerful than simply using the `naive' approximation $\bar{P}_{lam} = l/N$.

Below we describe how one can obtain an approximation of the confidence bands shown in figure \ref{fig:p_lam_subsampling}(b, d) based on one sample.
We first draw a random sample of $N = 10$ RPs for each of the $40$ energy levels, then compute the $40$ resulting posterior distributions using equation (\ref{eq:posterior}).
We then use this single sample to build a distribution of $1000$ fitting functions as follows.
From each of the posterior distributions, we draw a random laminarization probability.
The resulting values are fitted using the usual fitting function in equation (\ref{eq:fittingfunc}).
This procedure is the same as that described in figure \ref{fig:estimation_schema} with only one exception: instead of computing point estimates at step 3, we draw random values of the laminarization probabilities from the posterior distributions obtained at step 2. 
This is repeated $1000$ times, where each time $40$ new laminarization probabilities are drawn from the $40$ fixed posterior distributions. 
In this way, we build a distribution of fitting functions. 
Having calculated the distribution of fitting functions, we can readily compute its interquartile and interdecile ranges whose examples are shown in the left plots in figure \ref{fig:p_lam_subsampling} for the uncontrolled and controlled ($W_{osc} = 0.3, \omega = 1/16$) cases.
Although they are constructed using a single sample, they can be compared with the the results obtained on the whole dataset and represented on the right panels;
the width of the confidence bands obtained through Bayesian estimation (plots \ref{fig:p_lam_subsampling}(a, c)) is both qualitatively and quantitatively similar to that inferred from subsampling of a large database of RPs and thus treated as an accurate estimation (plots \ref{fig:p_lam_subsampling}(b, d)).
The most significant difference can be found for intermediate values of the energy in the controlled case (plots \ref{fig:p_lam_subsampling}(c, d)) where the confidence band resulting from Bayesian estimation is wider than the accurate counterpart.
In the same plots, we display the estimations of uncertainty levels of $P_{lam}(E_j)$ via the interquartile ranges: these are smaller for extreme values of the laminarization probability and larger for intermediate ones.
Despite the fact that these results highly depend on a particular choice of the initial sample, they show that Bayesian estimation is capable of giving a reasonable approximation of the confidence bands.

In addition to interdecile and interquartile ranges of the distribution of fitting functions, we can readily compute the confidence intervals for all the scalar metrics $S$, $E_{flex}$ and $E_a$ following exactly the same strategy.

\section{Control assessment via the laminarization probability}

\begin{figure}
    \includegraphics[width=1.0\textwidth]{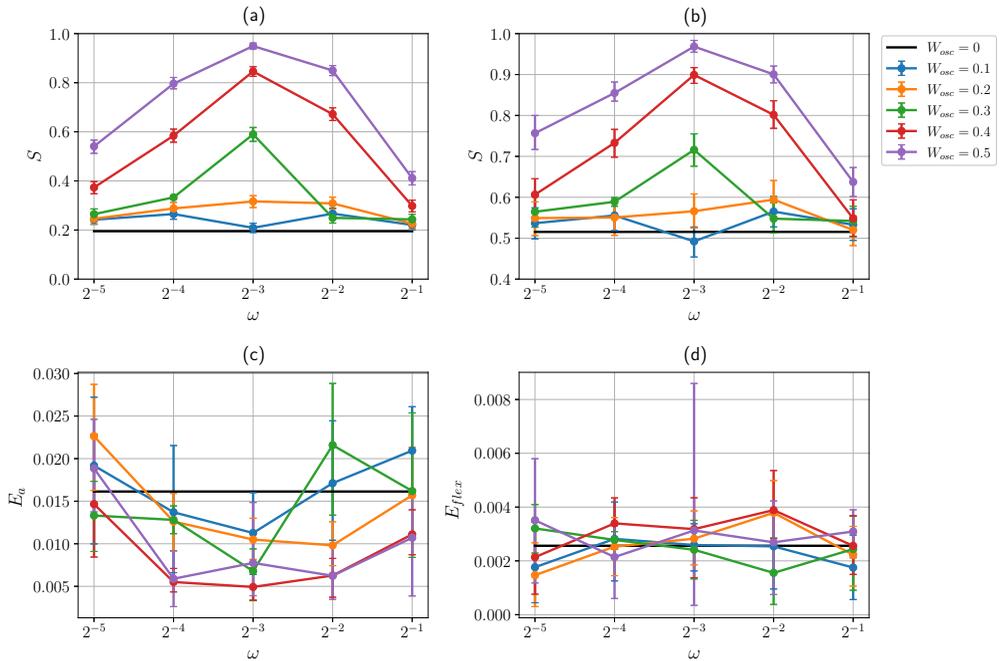}
    \caption{Performance of the wall oscillation control strategy as a function of the oscillation frequency $\omega$ shown via the laminarization score $S$ for the uniform (a) and the exponential (b) distributions of the RP energy, $f_E(E)$, and via energies $E_{a}$ (c) and $E_{flex}$ (d). The results are shown in each panel for different forcing amplitudes as indicated by the legend. Error bars show the confidence intervals (interdecile ranges) resulting from Bayesian estimation. The case $W_{osc} = 0.5$ and $\omega = 1/8$ leads to almost complete laminarization of all RPs ($p(E) \approx 1$), so that $p(E)$ only weakly depends on $E$, thereby explaining the abnormal variation of $E_{flex}$.}
    \label{fig:full_estimation}
\end{figure}

Using the method described above, we can now characterize the robustness of the laminar flow and find the optimal control parameter values (amplitude $W_{osc}$ and frequency $\omega$) by seeking to optimize the laminarization score $S$.
We estimate the dependence of $S$ and the associated confidence intervals on the control parameters by using $N=10$ RPs per energy level and then performing Bayesian estimation for $W_{osc} \in \{0.1, 0.2, \dots, 0.5\}$ and $\omega \in \{2^{-1}, 2^{-2}, \dots, 2^{-5}\}$.
To characterise the dependence of the fitting function on the amplitude and frequency of the wall oscillations, we additionally track the values of $E_a$ and $E_{flex}$.
The results are shown in figure \ref{fig:full_estimation}, where the uniform perturbation energy distribution $f^{(uni)}_E(E)$ is used to calculate $S$ in panel (a) and the exponential distribution $f^{(exp)}_E(E)$ is used in panel (b).
The most important observation is that in both plots, most of the confidence intervals obtained for $S$ do not overlap, so that we can clearly establish a hierarchy between the various control parameter values tested.

Independently of the type of distribution chosen for the RP energy, $S$ displays two trends as a function of $W_{osc}$ and $\omega$.
Firstly, we note that $S$ grows with respect to $W_{osc}$ except for a single case.
This is not the only flow for which increasing the amplitude has a favorable effect on the control performance: for example, studies of turbulent channel and pipe flows reported an increase in drag reduction when the oscillation amplitude was increased \citep{Quadrio2000, Quadrio2004}.
Secondly, $S$ reaches a maximum at $\omega = 1/8$ for all sufficiently large values of $W_{osc}$, and monotonically decays away from it.
These trends imply that wall oscillations make the laminar flow the most robust for $W_{osc} = 0.5$ and $\omega = 1/8$ for the range of control parameter values considered.
At these values of the control parameters, laminarization is nearly inevitable (only $3$ out of $400$ RPs were observed to transition to turbulence, yielding $S \approx 0.95$).
In fact, the control strategy is so efficient for $W_{osc} = 0.5$ that the concept of the laminarization probability becomes ill-posed since turbulence can rarely be sustained and so the separation between laminarizing and turbulent RPs becomes ambiguous.
All these observations regarding $S$ are valid for both the uniform and exponential distributions of RP energy.

Whilst the energy associated with the inflection point in the laminarization probability, $E_{flex}$, fluctuates without any clear dependence on $W_{osc}$ or $\omega$, the expected values of $E_a$ indicate that the beginning of the asymptotic regime of the laminarization probability seems to be negatively correlated with $S$ and reaches its minimum value at $W_{osc} = 0.5$ and $\omega = 1/8$.
These observations suggest that efficient control tends to increase the asymptotic value of the laminarization probability $a$.
However, as noted earlier, estimates for $E_a$ and $E_{flex}$ cannot be considered as reliable owing to the large confidence intervals associated with them.

Since the laminarization score determines how likely it is to observe laminar flow given a random initial condition drawn from a specified distribution, it is useful to consider the expected consumed energy assessed via the \textit{expected dissipation rate}:
\begin{equation}
  \bar{\epsilon} = S \epsilon_{lam}(W_{osc}, \omega) + (1 - S) \epsilon_{turb}(W_{osc}, \omega),
\end{equation}
where we recall that $\epsilon_{lam}$ and $\epsilon_{turb}$ are the dissipation rates of the laminar and turbulent flows given in equations (\ref{eq:lam_diss_rate}) and (\ref{eq:diss_rate}) respectively.
The expected dissipation rate can be thought of as a cost function which a flow control designer may seek to minimize with respect to the control parameters $W_{osc}$ and $\omega$.

\begin{figure}
    \includegraphics[width=1.0\textwidth]{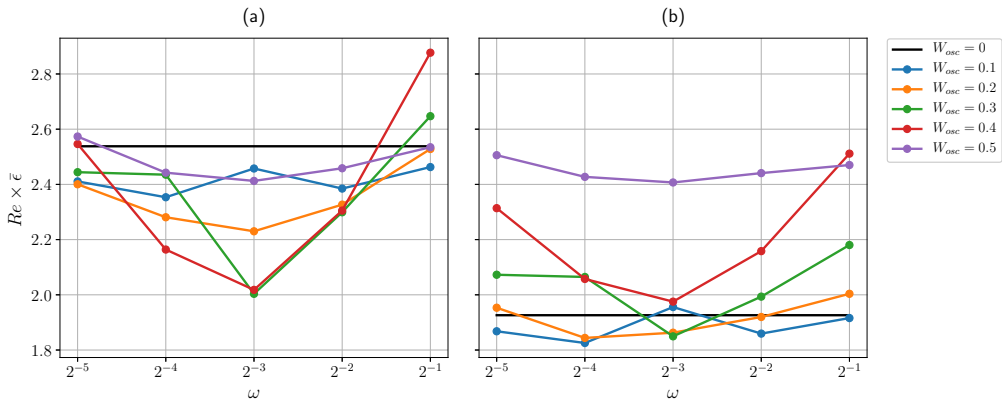}
    \caption{Expected dissipation rate for the uniform (a) and exponential (b) distributions of the RP energy plotted as functions of $W_{osc}$ and $\omega$.}
    \label{fig:expected_diss_rate}
\end{figure}

The dependence of the expected dissipation rate on the control parameter values is shown in figure \ref{fig:expected_diss_rate} for the laminarization score calculated with respect to the uniform (panel \ref{fig:expected_diss_rate}(a)) and exponential (panel \ref{fig:expected_diss_rate}(b)) distributions of the kinetic energy of RPs.
Since there is no data for the turbulent dissipation rate at $W_{osc} = 0.5$ (see the caption of figure \ref{fig:turb_prof_and_diss} for the explanation), we assume that $\epsilon_{turb} \approx 2.7$ which is equal to the smallest value among all the values of the turbulent dissipation rate shown in figure \ref{fig:turb_prof_and_diss}.
For a uniform distribution of RP energies, the wall oscillation at $W_{osc} = 0.3$ and $\omega = 1/8$ is found to minimize the expected dissipation rate, reducing it by approximately $20\%$ compared to the uncontrolled flow.
Increasing the control amplitude to $W_{osc} = 0.4$ and keeping the same control frequency performs nearly as well.
These results point toward a similar optimal control parameter value as \cite{Rabin2014} who obtained $W_{osc} = 0.35$ as the amplitude that maximizes the minimal seed energy at a different Reynolds number ($\Rey = 1000$).
The picture is different for an exponential distribution of RP energies: figure \ref{fig:expected_diss_rate}(b) shows that there are five combinations of the amplitude and frequency yielding an almost equally small value of the expected dissipation rate among which $W_{osc} = 0.1$ and $\omega = 1/16$ or $W_{osc} = 0.2$ and $\omega = 1/16$ are the two best options and decrease the energetic cost of the flow by $5\%$ compared to the uncontrolled case.
In comparison to the case of the uniform distribution, the reduction in the expected dissipation rate is not so pronounced for the exponential distribution.
Moreover, the vast majority of control configurations for this case actually increase the expected dissipation rate (the corresponding control parameter values lie above the black line in figure \ref{fig:expected_diss_rate}(b)).
The lack of performance of the wall-oscillation strategy is not surprising in the case of exponentially distributed perturbations: this type of control was shown to mostly increase the laminarization probability for large perturbation energies \citep{Pershin2020} and these are less likely to be generated in the exponential case than for the uniform case.
The appropriate choice of the amplitude and frequency consequently depends on the distribution of the perturbation energy prescribed by the problem at hand.
Moreover, it is also possible that in certain cases, the expected dissipation rate is not the appropriate choice for the cost function.
For example, a flow control designer may conclude that transition to turbulence is so harmful (e.g., due to extreme events occurring in turbulent flows) that it is better to pick $W_{osc} = 0.5$ and $\omega = 1/8$ to suppress transition entirely despite the additional cost this choice incurs.

\section{Discussion}

In this work, we have studied the nonlinear stability of plane Couette flow under the action of a well-known control strategy: in-phase wall oscillations in the spanwise direction.
We first briefly characterized the typical turbulence observed in this flow under the action of the control strategy and observed that it becomes less energetic as the oscillation amplitude is increased.
We also found that the turbulent kinetic energy is at its lowest for oscillation frequencies around $\omega = 1/16$.
We then turned to the two most studied scalar indicators of the robustness of the laminar flow: the turbulent kinetic energy of the edge state and that of the minimal seed.
The edge state results proved misleading: their energy typically decreases under the effect of the control strategy, which could lead to the erroneous conclusion that the laminar flow has become more sensitive to perturbations.
The reason for this tempting misinterpretation is topological: edge states are attractors on the edge of chaos but they do not have to be representative of any of its properties.
The minimal seed, by definition, is the minimal energy perturbation of the laminar flow that transitions to turbulence.
Under the effect of our control, its turbulent kinetic energy increases, indicating that triggering turbulence requires larger amplitude perturbations, as was observed by \citet{Rabin2014} at a different value of the Reynolds number.
In spite of this, our calculations showed that the minimal seed turbulent kinetic energy is much smaller than the values that the turbulent kinetic energy typically takes along the edge of chaos and, as a result, it is of limited use for practical control design.
The trajectory of the minimal seed, however, provides a great deal of information regarding the processes at play during transition, which, in turn, can be controlled \citep{Pringle2012, Duguet2013, Cherubini2013, Kerswell2018}. 
This is, however, out of the scope of our investigation.

To appreciate the structure of the edge of chaos in more detail, we turned to the laminarization probability, a concept recently developed \citep{Pershin2020} which represents the probability that a perturbation of the laminar flow decays as a function of its turbulent kinetic energy.
The laminarization probability can be understood as a measure of the relative volume in state space of the basin of attraction of the laminar flow.
Obtaining converged laminarization probability results can be computationally prohibitive: \citet{Pershin2020} required $31200$ core-hours on state-of-the-art computational facilities ($7800$ simulations distributed onto $39$ energy levels, each simulation lasting approximately $4$ core-hours) to accurately determine a single laminarization probability curve.
We developed an efficient Bayesian approach that allows us to reduce the computation time of laminarization probability curves by $95\%$.
We used this approach to compute the laminarization probability for a number of control amplitudes and frequencies and determined that the laminar flow becomes more robust to finite-amplitude perturbations as the control amplitude is increased and as the control frequency approaches $\omega = 1/8$.
Interestingly, the same value of the frequency, $\omega = 1/8$, was found to minimize the turbulent kinetic energy (see figure \ref{fig:turb_attractor_estimate}) and the wall-normal shear rate variation (see figure \ref{fig:turb_prof_and_diss}(a)), but held no importance for the edge state or minimal seed energies.

To provide a simple, explicit assessment of the performance of control strategies, we further introduced two quantities based on the laminarization probability: the laminarization score and the expected dissipation rate.
The former represents the expected laminarization probability given that random perturbations are drawn with a prescribed turbulent kinetic energy distribution and is the quantity that needs to be maximized in applications where turbulence should be suppressed at all costs.
The latter computes the expected energetic cost of the flow taking into account occurrences of laminar and turbulent regimes with their respective probabilities and is, as such, to be minimized in applications looking for energetic efficiency.
We tested these two scalar quantities for different distributions of perturbation energies and obtained a clear hierarchy of control parameter values, in terms of the control performance.
The laminarization score confirmed the effect of the amplitude and frequency of wall oscillations on the laminar flow stability. 
On the other hand, the expected dissipation rate allowed us to determine the control parameters for which more energy is spent maintaining the wall oscillations than is saved by the increased laminarization probability.

We believe that our probabilistic protocol based on the Bayesian estimation of the laminarization probability provides an efficient approach to reliably assess the global robustness of the laminar flow.
In contrast to minimal-seed or edge-state approaches, our approach provides overall information on the structure of the edge of chaos.
It simply relies on the use of a time-stepper and does not require complex optimization algorithms, which makes its implementation relatively user-friendly.
Finally, the laminarization probability can be utilized to construct a control assessment framework and determine optimal control conditions, as demonstrated in this paper.
As formulated, the only pre-requisite is the knowledge of the form of perturbations that the flow is subject to and their probability as a function of their amplitude.

Our method is by no means limited to plane Couette flow and the particular control strategy we considered.
We expect it to be applicable to the study and control of any nonlinear system exhibiting finite-amplitude instability \citep{Xi2012, Zammert2015, Watanabe2016, Chantry2017, Khan2021}.

\section*{Acknowledgments}

AP is grateful for the support of Professor Timothy Palmer and the European Research Council grant ITHACA (Grant Agreement No. 741112) at the University of Oxford.
CB acknowledges support from the Leverhulme Trust under Research Project Grant RPG-2018-311.
SMT is supported by the European Research Council (ERC) under the European Union Horizon
2020 research and innovation program (Grant Agreement No.
D5S-DLV-786780) and in part
by a grant from the Simons Foundation (Grant No. 662962, GF).
This work was undertaken on ARC3 and ARC4, part of the High Performance Computing facilities at the University of Leeds, UK.

\section*{Declaration of Interests}
The authors report no conflict of interest.

\appendix
\section{Laminar solution for plane Couette flow with in-phase spanwise wall oscillations}\label{app:deriv_osc_laminar}

Consider the Navier--Stokes equation for an incompressible flow:
\begin{eqnarray}
\label{NSE1}
\partial_t \boldsymbol{U} + (\boldsymbol{U} \cdot \nabla) \boldsymbol{U} &=& -\nabla P + \frac{1}{\Rey} \nabla^2 \boldsymbol{U}, \\
\label{NSE2}
\nabla \cdot \boldsymbol{U} &=& 0,
\end{eqnarray}
where $\boldsymbol{U} = (U,V,W)$ is the velocity field of components $U$, $V$ and $W$ in the streamwise ($x$), wall-normal ($y$) and spanwise ($z$) directions respectively, and where $P$ is the pressure and $\Rey$ is the Reynolds number.
This system is accompanied with boundary conditions associated with in-phase spanwise wall oscillation:
\begin{equation}
\boldsymbol{U}(x, \pm 1, z, t) = \left[\pm 1, 0, W_{osc} \sin(\omega t + \phi)\right],
\end{equation}
where $W_{osc}$, $\omega$ and $\phi$ are the amplitude, frequency and phase of the wall oscillations.
The problem is closed by the imposition of periodic boundary conditions in the $x-$ and $z-$directions.

Consider a solution of the form:
\begin{equation}
\boldsymbol{U}(x, y, z, t) = [U(y), 0, W(y, t)].
\end{equation}
After substituting into system (\ref{NSE1}), (\ref{NSE2}) and assuming constant pressure, we get
\begin{eqnarray}
\partial_{yy} U &=& 0, \\
\partial_t W &=& \frac{1}{\Rey} \partial_{yy} W. \label{eq:W}
\end{eqnarray}
The equation for $U$ yields the laminar solution for plane Couette flow, $U(y) = y$, whilst the equation for $W$ corresponds to a diffusion equation with boundary condition $W(\pm 1, t) = W_{osc} \sin(\omega t + \phi)$.
To find the solution for $W$, we proceed to the following change of variable:
\begin{equation}
g(y, t) = W(y, t) - W_{osc} \sin(\omega t + \phi),
\end{equation}
satisfying homogeneous Dirichlet boundary conditions which, after substitution into (\ref{eq:W}), yields the inhomogeneous diffusion equation:
\begin{equation}
\label{eq:g_eq}
\partial_t g = \frac{1}{\Rey} \partial_{yy} g - W_{osc} \omega \cos(\omega t + \phi).
\end{equation}
We first consider the homogeneous version of this equation.
Separating the variables into $g(y, t) = \theta_k(t) \psi_k(y)$ leads to two eigenvalues problems involving a constant $\lambda$:
\begin{eqnarray}
  \label{thetak}
\frac{\mathrm{d}\theta_k}{\mathrm{d} t} &=& \lambda \theta_k, \\
\frac{\mathrm{d}^2 \psi_k}{\mathrm{d} y^2} &=& \Rey \lambda \psi_k.
\end{eqnarray}
Given homogeneous boundary conditions, we get:
\begin{equation}
\psi_k(y) = \sin\left[\frac{1}{2}k \pi (y + 1)\right],
\end{equation}
where $k \pi = \sqrt{-\Rey \lambda}$, $k \in \mathbb{N}$.

Instead of solving equation (\ref{thetak}) directly, we can project equation (\ref{eq:g_eq}) onto the following eigenmodes:
\begin{equation}
\label{eq:g_decomp}
g(y, t) = \sum\limits_{\substack{k = 1,\\ k \in \mathbb{N}}}^{\infty} A_k(t) \sin\left[\frac{1}{2}k \pi (y + 1)\right]
\end{equation}
and write its rightmost term as:
\begin{equation}
\label{eq:F_decomp}
- W_{osc} \omega \cos(\omega t + \phi) \sum\limits_{\substack{k = 1,\\ k \in \mathbb{N}}}^{\infty} \frac{2 \left[1 + (-1)^{k+1}\right]}{k\pi} \sin\left[\frac{1}{2}k \pi (y + 1)\right].
\end{equation}
Substituting (\ref{eq:g_decomp}) and (\ref{eq:F_decomp}) into (\ref{eq:g_eq}) and collecting the terms proportional to $\sin\left[k \pi (y + 1) / 2\right]$ yields a first-order ODE for $A_k(t)$:
\begin{equation}
\frac{\mathrm{d} A_k}{\mathrm{d} t} + \frac{k^2 \pi^2}{\Rey} A_k = - \frac{2 \left[1 + (-1)^{k+1}\right]}{k\pi} W_{osc} \omega \cos(\omega t + \phi).
\end{equation}
It has general solution:
\begin{equation}
A_k = \alpha_{k} \exp\left[-\frac{k^2 \pi^2}{\Rey} t\right] + a_{k} \sin(\omega t + \phi) + b_{k} \cos(\omega t + \phi),
\end{equation}
where $\alpha_{k}$, $a_{k}$ and $b_{k}$ are constants to be determined.
Since we are looking for a laminar flow solution, the transient term can be dropped by letting $t \to \infty$, which yields a time-periodic solution for $A_k$.

Thus, in a general form, the solution for $g(y, t)$ can be written as follows:
\begin{equation}
\label{eq:g_ansatz}
g(y, t) = a(y) \sin(\omega t + \phi) + b(y) \cos(\omega t + \phi),
\end{equation}
where the functions $a(y)$ and $b(y)$ are infinite sine series and must both satisfy homogeneous Dirichlet boundary conditions.
Substitution into (\ref{eq:g_eq}) yields:
\begin{multline}
\omega a \cos(\omega t + \phi) - \omega b \sin(\omega t + \phi) \\= \frac{a^{\prime\prime}}{\Rey} \sin(\omega t + \phi) + \frac{b^{\prime\prime}}{\Rey} \cos(\omega t + \phi) - W_{osc} \omega \cos(\omega t + \phi),
\end{multline}
where primes denote derivatives with respect to $y$. 
Collecting terms proportional to the $\sin$ and $\cos$ bases gives the following system:
\begin{eqnarray}
b &=& -\frac{a^{\prime\prime}}{\omega \Rey}, \label{eq:b_ode} \\
a^{\prime\prime\prime\prime} + \omega^2 \Rey^2 a &=& - W_{osc} \omega^2 \Rey^2, \label{eq:a_ode}
\end{eqnarray}
where the fourth-order equation for $a(y)$ is accompanied with the following boundary conditions:
\begin{eqnarray}
a(-1) &=& a(1) = 0, \\
a^{\prime\prime}(-1) &=& a^{\prime\prime}(1) = 0.
\end{eqnarray}
The general solution for real-valued $a(y)$ has the following form:
\begin{multline}
\label{eq:a_shifted}
a(y) = a_1 e^{\Omega (y-1)} \cos\Omega (y-1) + a_2 e^{\Omega (y-1)} \sin\Omega (y-1) \\+ a_3 e^{-\Omega (y-1)} \cos\Omega (y-1) + a_4 e^{-\Omega (y-1)} \sin\Omega (y-1) - W_{osc}.
\end{multline}
where $(a_1, a_2, a_3, a_4) \in \mathbb{R}^4$ and $\Omega = \sqrt{\omega \Rey/2}$.

Equation (\ref{eq:b_ode}) then gives the general solution for $b(y)$:
\begin{multline}
\label{eq:b_shifted}
b(y) = a_1 e^{\Omega (y-1)} \sin\Omega (y-1) - a_2 e^{\Omega (y-1)} \cos\Omega (y-1) \\- a_3 e^{-\Omega (y-1)} \sin\Omega (y-1) + a_4 e^{-\Omega (y-1)} \cos\Omega (y-1).
\end{multline}
Boundary conditions $a(1) = b(1) = 0$ immediately give $a_3 = W_{osc} - a_1$ and $a_4 = a_2$, whereas boundary conditions $a(-1) = b(-1) = 0$ give the following system with respect to $a_1$ and $a_2$:
\begin{equation}
\begin{bmatrix}
2 \sinh 2\Omega \cos 2\Omega && 2 \cosh 2\Omega \sin 2\Omega \\
2 \cosh 2\Omega \sin 2\Omega && -2 \sinh 2\Omega \cos 2\Omega
\end{bmatrix}
\begin{bmatrix}
a_1 \\
a_2
\end{bmatrix}
=
\begin{bmatrix}
W_{osc}\left( e^{2\Omega} \cos 2\Omega - 1 \right) \\
W_{osc} e^{2\Omega} \sin 2\Omega
\end{bmatrix},
\end{equation}
whose solution is
\begin{eqnarray}
a_1 &=& \frac{W_{osc} \left( e^{2\Omega} + \cos 2\Omega \right)}{2 (\cosh 2\Omega + \cos 2\Omega)}, \\
a_2 &=& -\frac{W_{osc} \sin 2\Omega}{2 (\cosh 2\Omega + \cos 2\Omega)}.
\end{eqnarray}
Finally, substituting the expressions for $a_1, a_2, a_3, a_4$ into (\ref{eq:a_shifted}) and (\ref{eq:b_shifted}) and then $a(y)$ and $b(y)$ into $g(y, t)$ yields the solution for $W(y, t)$:
\begin{eqnarray}
W(y, t) = &\displaystyle\frac{W_{osc}}{\cosh 2\Omega + \cos 2\Omega} \Big[ \left[ \cosh\Omega y_+ \cos\Omega y_- + \cosh\Omega y_- \cos\Omega y_+ \right] \sin(\omega t + \phi) \nonumber \\ 
& + \left[ \sinh\Omega y_+ \sin\Omega y_- + \sinh\Omega y_- \sin\Omega y_+ \right] \cos(\omega t + \phi) \Big],
\end{eqnarray}
where $y_{\pm} = y \pm 1$.

\section{Uninformative priors in Bayesian inference}\label{app:priors}

Uninformative priors are supposed to contain as little information about the considered parameter ($P_{lam}$ in our case) as possible.
There exist several ``principles'' to help determine which prior distribution should be used.
In this work, we consider two of them: the principle of maximum entropy \citep{Jaynes1957} and that of reference priors \citep{Bernardo1979}.

The principle of maximum entropy, suggests that the prior distribution must maximize the uncertainty or, equivalently, maximize the surprisal.
For continuous distributions, the uncertainty can be quantified by the \textit{differential entropy}:
\begin{equation}
\label{eq:diff_entropy}
h[f_{P_{lam}}] = -\int_{0}^{1} f_{P_{lam}}(p)\log_2 f_{P_{lam}}(p) \mathrm{d}p,
\end{equation}
which is the analogue of the Shannon information entropy \citep{Shannon1948}.
In the absence of constraints, the distribution with finite support maximizing functional (\ref{eq:diff_entropy}) is the uniform prior distribution $f_{P_{lam}}(p) = 1$ \citep{Park2009}.

The second principle, i.e. the approach based on choosing reference priors, implies constructing a prior distribution that maximizes the amount of information about $P_{lam}$ gained after observing data $\boldsymbol{r}$ \citep{Bernardo1979}.
The functional to be maximized in this approach is the expected Kullback--Leibler divergence of the prior distribution $f_{P_{lam}}(p)$ from the posterior distribution $f_{P_{lam}}(p | \boldsymbol{R} = \boldsymbol{r})$:
\begin{equation}
\label{eq:mutual_information}
I[f_{P_{lam}}] = \mathbb{E} D(f_{P_{lam}}(p | \boldsymbol{R} = \boldsymbol{r}) || f_{P_{lam}}(p)),
\end{equation} 
where $\mathbb{E}$ denotes the expected value and the \textit{Kullback--Leibler divergence}, $D(\cdot \ || \ \cdot)$, is defined as follows:
\begin{equation}
D(f_{P_{lam}}(p | \boldsymbol{R} = \boldsymbol{r}) \ || \ f_{P_{lam}}(p)) = \int_0^1 f_{P_{lam}}(p | \boldsymbol{R} = \boldsymbol{r}) \log_2 \frac{f_{P_{lam}}(p | \boldsymbol{R} = \boldsymbol{r})}{f_{P_{lam}}(p)} \mathrm{d}p.
\end{equation}
This quantity can be understood as a measure of the information about $P_{lam}$ gained by replacing the prior distribution with the posterior one where the latter was obtained after observing sample data $\boldsymbol{r}$.
To eliminate its dependence on the sample $\boldsymbol{R}$ whose particular value $\boldsymbol{r}$ is unknown for us \emph{a priori}, we take the expectation of the Kullback--Leibler divergence $D(f_{P_{lam}}(p | \boldsymbol{R} = \boldsymbol{r}) || f_{P_{lam}}(p))$ with respect to $\boldsymbol{R}$ which gives the expression (\ref{eq:mutual_information}).
Its maximization with respect to $f_{P_{lam}}(p)$ can then be thought of as looking for a prior distribution that, on average, maximizes the ``degree of surprisal'' associated with $P_{lam}$ after observing sample data $\boldsymbol{r}$.
A class of such uninformative priors is referred to as \textit{reference priors}.
In the case of our likelihood function (\ref{eq:likelihood}), the resulting prior is
\begin{equation}
f_{P_{lam}}(p) = \frac{1}{\pi \sqrt{p(1-p)}},
\end{equation}
which is also known to belong to the class of \textit{Jeffreys' priors} which originates from the use of the principle of transformation groups \citep{Box2011}.

\section{Generation of random perturbations}\label{app:rp_generation}

Random perturbations (RPs) used in this study were introduced by \cite{Pershin2020} and are defined as follows
\begin{equation}
\label{eq:rp_def}
\boldsymbol{u} = A \boldsymbol{u}_{\perp} + B \boldsymbol{U_{\text{lam}}},
\end{equation}
where $\boldsymbol{u}_{\perp}$ is a random velocity field, $\boldsymbol{U_{\text{lam}}}$ is the laminar solution of plane Couette flow in the absence of any control and $A$ and $B$ are random numbers.
The random velocity field $\boldsymbol{u}_{\perp}$ is incompressible and satisfies no-slip boundary conditions.
It is generated using subroutine \textit{randomfield} from \textit{Channeflow} \citep{Gibson2014channelflow} by drawing its spectral coefficients $(\hat{u},\hat{v},\hat{w})_{ijk}$, where $i$, $j$, $k$ correspond to the streamwise, wall-normal and spanwise wavenumbers respectively, from the uniform distribution and then scaling them so that spectral coefficients decay exponentially with respect to the size of the wavenumber vector:
\begin{equation}
(\hat{u},\hat{v},\hat{w})_{ijk} = (Z_u,Z_v,Z_w) \times (1 - \sigma)^{|i| + |j| + |k|},
\end{equation}
where $(Z_u,Z_v,Z_w)$ is a triplet of random numbers drawn from the uniform distribution with support $[-1; 1]$ and $\sigma = 0.4$ is a decay parameter.
Once the spectral coefficients are generated, the random component $\boldsymbol{u}_{\perp}$ is corrected to ensure incompressibility and no-slip boundary conditions.
Additionally, $\boldsymbol{u}_{\perp}$ is made orthogonal to the laminar solution, i.e., $\langle \boldsymbol{u}_{\perp}, \boldsymbol{U_{\text{lam}}} \rangle = 0$, and normalized so that $||\boldsymbol{u}_{\perp}||^2 = 1$.
Next, coefficients $A$ and $B$ from \eqref{eq:rp_def} are generated so that the RP has a prescribed value of the kinetic energy $E$.
It is done by drawing $B$ from the uniform distribution with support $[-2E / ||\boldsymbol{U_{\text{lam}}}||; 2E / ||\boldsymbol{U_{\text{lam}}}||]$ and then computing $A = \sqrt{2E - B^2 ||\boldsymbol{U_{\text{lam}}}||}$.
Finally, we time-integrate the resulting field $\boldsymbol{u}$ for two tiny time steps to ensure no-slip boundary conditions which introduces only a negligibly small change in the kinetic energy of the RP and in the values of $A$ and $B$.


\bibliographystyle{jfm}
\bibliography{2020_Name}

\begin{thebibliography}{44}
\expandafter\ifx\csname natexlab\endcsname\relax\def\natexlab#1{#1}\fi
\def\au#1{#1} \def\ed#1{#1} \def\yr#1{#1}\def\at#1{#1}\def\jt#1{\textit{#1}}
  \def\bt#1{#1}\def\bvol#1{\textbf{#1}} \def\vol#1{#1} \def\pg#1{#1}
  \def\publ#1{#1}\def\arxiv#1{#1}\def\org#1{#1}\def\st#1{\textit{#1}}

\bibitem[Barkley(2016)]{Barkley2016}
{\sc \au{Barkley, D.}} \yr{2016}  \at{Theoretical perspective on the route to
  turbulence in a pipe}.  \jt{J. Fluid Mech.}  \bvol{803},  \pg{P1}.

\bibitem[Bernardo(1979)]{Bernardo1979}
{\sc \au{Bernardo, J.~M.}} \yr{1979}  \at{Reference posterior distributions for
  {B}ayesian inference}.  \jt{J. R. Stat. Soc. Ser. B Stat. Methodol.}
  \bvol{41}~(2),  \pg{113--128}.

\bibitem[Box \& Tiao(2011)]{Box2011}
{\sc \au{Box, G. E.~P.} \& \au{Tiao, G.~C.}} \yr{2011} {\em Bayesian inference
  in statistical analysis\/}, ,  \vol{vol.~40}.  \publ{John Wiley \& Sons}.

\bibitem[Budanur {\em et~al.\/}(2020)Budanur, Marensi, Willis \&
  Hof]{Budanur2020}
{\sc \au{Budanur, N.~B.}, \au{Marensi, E.}, \au{Willis, A.~P.} \& \au{Hof, B.}}
  \yr{2020}  \at{Upper edge of chaos and the energetics of transition in pipe
  flow}.  \jt{Phys. Rev. Fluids}  \bvol{5}~(2),  \pg{023903}.

\bibitem[Chantry \& Schneider(2014)]{Chantry2014}
{\sc \au{Chantry, M.} \& \au{Schneider, T.~M.}} \yr{2014}  \at{Studying edge
  geometry in transiently turbulent shear flows}.  \jt{J. Fluid Mech.}
  \bvol{747},  \pg{506--517}.

\bibitem[Chantry {\em et~al.\/}(2017)Chantry, Tuckerman \&
  Barkley]{Chantry2017}
{\sc \au{Chantry, M.}, \au{Tuckerman, L.~S.} \& \au{Barkley, D.}} \yr{2017}
  \at{Universal continuous transition to turbulence in a planar shear flow}.
  \jt{J. Fluid Mech.}  \bvol{824}.

\bibitem[Cherubini \& {De Palma}(2013)]{Cherubini2013}
{\sc \au{Cherubini, S.} \& \au{{De Palma}, P.}} \yr{2013}  \at{{Nonlinear
  optimal perturbations in a {C}ouette flow: bursting and transition}}.  \jt{J.
  Fluid Mech.}  \bvol{716},  \pg{251--279}.

\bibitem[Cherubini {\em et~al.\/}(2011)Cherubini, Palma, Robinet \&
  Bottaro]{Cherubini2011}
{\sc \au{Cherubini, S.}, \au{Palma, P.~D.}, \au{Robinet, J.-Ch.} \&
  \au{Bottaro, A.}} \yr{2011}  \at{Edge states in a boundary layer}.  \jt{Phys.
  Fluids}  \bvol{23}~(5),  \pg{051705}.

\bibitem[Couliou \& Monchaux(2015)]{Couliou2015}
{\sc \au{Couliou, M.} \& \au{Monchaux, R.}} \yr{2015}  \at{Large-scale flows in
  transitional plane {C}ouette flow: a key ingredient of the spot growth
  mechanism}.  \jt{Phys. Fluids}  \bvol{27}~(3),  \pg{034101}.

\bibitem[Dauchot \& Daviaud(1995)]{Dauchot1995}
{\sc \au{Dauchot, O.} \& \au{Daviaud, F.}} \yr{1995}  \at{Finite amplitude
  perturbation and spots growth mechanism in plane {C}ouette flow}.  \jt{Phys.
  Fluids}  \bvol{7}~(2),  \pg{335--343}.

\bibitem[Duguet {\em et~al.\/}(2013)Duguet, Monokrousos, Brandt \&
  Henningson]{Duguet2013}
{\sc \au{Duguet, Y.}, \au{Monokrousos, A.}, \au{Brandt, L.} \& \au{Henningson,
  S.}} \yr{2013}  \at{Minimal transition thresholds in plane {C}ouette flow}.
  \jt{Phys. Fluids}  \bvol{25},  \pg{084103}.

\bibitem[Duguet {\em et~al.\/}(2009)Duguet, Schlatter \&
  Henningson]{Duguet2009}
{\sc \au{Duguet, Y.}, \au{Schlatter, P.} \& \au{Henningson, D.~S.}} \yr{2009}
  \at{Localized edge states in plane {C}ouette flow}.  \jt{Phys. Fluids}
  \bvol{21}~(11),  \pg{111701}.

\bibitem[Duguet {\em et~al.\/}(2010)Duguet, Schlatter \&
  Henningson]{Duguet2010}
{\sc \au{Duguet, Y.}, \au{Schlatter, P.} \& \au{Henningson, D.~S.}} \yr{2010}
  \at{Formation of turbulent patterns near the onset of transition in plane
  {C}ouette flow}.  \jt{J. Fluid Mech.}  \bvol{650},  \pg{119}.

\bibitem[Eaves \& Caulfield(2015)]{Eaves2015}
{\sc \au{Eaves, T.~S.} \& \au{Caulfield, C.~P.}} \yr{2015}  \at{Disruption of
  {SSP/VWI} states by a stable stratification}.  \jt{J. Fluid Mech.}
  \bvol{784},  \pg{548--564}.

\bibitem[Gibson(2014)]{Gibson2014channelflow}
{\sc \au{Gibson, J.~F.}} \yr{2014}  \bt{{Channelflow}: {A} spectral
  {Navier--Stokes} simulator in {C}++}. {\em Tech. Rep.\/}.  \org{U. New
  Hampshire}, {\tt {Channelflow.org}}.

\bibitem[Jaynes(1957)]{Jaynes1957}
{\sc \au{Jaynes, E.~T.}} \yr{1957}  \at{Information theory and statistical
  mechanics}.  \jt{Phys. Rev.}  \bvol{106}~(4),  \pg{620}.

\bibitem[Kerswell(2018)]{Kerswell2018}
{\sc \au{Kerswell, R.~R.}} \yr{2018}  \at{Nonlinear nonmodal stability theory}.
   \jt{Annu. Rev. Fluid Mech.}  \bvol{50},  \pg{319--345}.

\bibitem[Khan {\em et~al.\/}(2021)Khan, Anwer, Hasan \& Sanghi]{Khan2021}
{\sc \au{Khan, H.~H.}, \au{Anwer, S.~F.}, \au{Hasan, N.} \& \au{Sanghi, S.}}
  \yr{2021}  \at{Laminar to turbulent transition in a finite length square duct
  subjected to inlet disturbance}.  \jt{Phys. Fluids}  \bvol{33}~(6),
  \pg{065128}.

\bibitem[Khapko {\em et~al.\/}(2016)Khapko, Kreilos, Schlatter, Duguet,
  Eckhardt \& Henningson]{Khapko2016}
{\sc \au{Khapko, T.}, \au{Kreilos, T.}, \au{Schlatter, P.}, \au{Duguet, Y.},
  \au{Eckhardt, B.} \& \au{Henningson, D.~S.}} \yr{2016}  \at{Edge states as
  mediators of bypass transition in boundary-layer flows}.  \jt{J. Fluid Mech.}
   \bvol{801},  \pg{R2}.

\bibitem[Kreilos {\em et~al.\/}(2016)Kreilos, Khapko, Schlatter, Duguet,
  Henningson \& Eckhardt]{Kreilos2016}
{\sc \au{Kreilos, T.}, \au{Khapko, T.}, \au{Schlatter, P.}, \au{Duguet, Y.},
  \au{Henningson, D.~S.} \& \au{Eckhardt, B.}} \yr{2016}  \at{Bypass transition
  and spot nucleation in boundary layers}.  \jt{Phys. Rev. Fluids}
  \bvol{1}~(4),  \pg{043602}.

\bibitem[Lecoanet \& Kerswell(2018)]{Lecoanet2018}
{\sc \au{Lecoanet, D.} \& \au{Kerswell, R.~R.}} \yr{2018}  \at{Connection
  between nonlinear energy optimization and instantons}.  \jt{Phys. Rev. E}
  \bvol{97},  \pg{012212}.

\bibitem[Meseguer \& Trefethen(2003)]{Meseguer2003}
{\sc \au{Meseguer, A.} \& \au{Trefethen, L.~N.}} \yr{2003}  \at{Linearized pipe
  flow to {R}eynolds number $10^7$}.  \jt{J. Comp. Phys.}  \bvol{186},
  \pg{178--197}.

\bibitem[Monokrousos {\em et~al.\/}(2011)Monokrousos, Bottaro, Brandt, {Di
  Vita} \& Henningson]{Monokrousos2011}
{\sc \au{Monokrousos, A.}, \au{Bottaro, A.}, \au{Brandt, L.}, \au{{Di Vita},
  A.} \& \au{Henningson, D.~S.}} \yr{2011}  \at{{Nonequilibrium thermodynamics
  and the optimal path to turbulence in shear flows}}.  \jt{Phys. Rev. Lett.}
  \bvol{106},  \pg{134502}.

\bibitem[Orszag(1971)]{Orszag1971}
{\sc \au{Orszag, S.~A.}} \yr{1971}  \at{Accurate solution of the
  {Orr--Sommerfeld} stability equation}.  \jt{J. Fluid Mech.}  \bvol{50},
  \pg{689--703}.

\bibitem[Park \& Bera(2009)]{Park2009}
{\sc \au{Park, S.~Y.} \& \au{Bera, A.~K.}} \yr{2009}  \at{Maximum entropy
  autoregressive conditional heteroskedasticity model}.  \jt{J. Econom.}
  \bvol{150}~(2),  \pg{219--230}.

\bibitem[Pershin {\em et~al.\/}(2020)Pershin, Beaume \& Tobias]{Pershin2020}
{\sc \au{Pershin, A.}, \au{Beaume, C.} \& \au{Tobias, S.~M.}} \yr{2020}  \at{A
  probabilistic protocol for the assessment of transition and control}.  \jt{J.
  Fluid Mech.}  \bvol{895},  \pg{A16}.

\bibitem[Pringle {\em et~al.\/}(2012)Pringle, Willis \& Kerswell]{Pringle2012}
{\sc \au{Pringle, C. C.~T.}, \au{Willis, A.~P.} \& \au{Kerswell, R.~R.}}
  \yr{2012}  \at{{Minimal seeds for shear flow turbulence: using nonlinear
  transient growth to touch the edge of chaos}}.  \jt{J. Fluid Mech.}
  \bvol{702},  \pg{415--443}.

\bibitem[Quadrio \& Ricco(2004)]{Quadrio2004}
{\sc \au{Quadrio, M.} \& \au{Ricco, P.}} \yr{2004}  \at{Critical assessment of
  turbulent drag reduction through spanwise wall oscillations}.  \jt{J. Fluid
  Mech.}  \bvol{521},  \pg{251--271}.

\bibitem[Quadrio \& Sibilla(2000)]{Quadrio2000}
{\sc \au{Quadrio, M.} \& \au{Sibilla, S.}} \yr{2000}  \at{Numerical simulation
  of turbulent flow in a pipe oscillating around its axis}.  \jt{J. Fluid
  Mech.}  \bvol{424},  \pg{217--241}.

\bibitem[Rabin {\em et~al.\/}(2012)Rabin, Caulfield \& Kerswell]{Rabin2012}
{\sc \au{Rabin, S. M.~E.}, \au{Caulfield, C.~P.} \& \au{Kerswell, R.~R.}}
  \yr{2012}  \at{{Triggering turbulence efficiently in plane {C}ouette flow}}.
  \jt{J. Fluid Mech.}  \bvol{712},  \pg{244--272}.

\bibitem[Rabin {\em et~al.\/}(2014)Rabin, Caulfield \& Kerswell]{Rabin2014}
{\sc \au{Rabin, S. M.~E.}, \au{Caulfield, C.~P.} \& \au{Kerswell, R.~R.}}
  \yr{2014}  \at{Designing a more nonlinearly stable laminar flow via boundary
  manipulation}.  \jt{J. Fluid Mech.}  \bvol{738},  \pg{R1}.

\bibitem[Rinaldi {\em et~al.\/}(2018)Rinaldi, Schlatter \&
  Bagheri]{Rinaldi2018}
{\sc \au{Rinaldi, E.}, \au{Schlatter, P.} \& \au{Bagheri, S.}} \yr{2018}
  \at{Edge state modulation by mean viscosity gradients}.  \jt{J. Fluid Mech.}
  \bvol{838},  \pg{379--403}.

\bibitem[Romanov(1973)]{Romanov1973}
{\sc \au{Romanov, V.~A.}} \yr{1973}  \at{Stability of plane-parallel {C}ouette
  flow}.  \jt{Funct. Anal. Appl.}  \bvol{7}~(2),  \pg{137--146}.

\bibitem[Schmid \& Henningson(2001)]{Schmid2001}
{\sc \au{Schmid, P.~J.} \& \au{Henningson, D.~S.}} \yr{2001} {\em Stability and
  transition in shear flows\/}, ,  \vol{vol. 142}.  \publ{Springer-Verlag New
  York, Inc.}

\bibitem[Schneider {\em et~al.\/}(2007)Schneider, Eckhardt \&
  Yorke]{Schneider2007}
{\sc \au{Schneider, T.~M.}, \au{Eckhardt, B.} \& \au{Yorke, J.~A.}} \yr{2007}
  \at{Turbulence transition and the edge of chaos in pipe flow}.  \jt{Phys.
  Rev. Lett.}  \bvol{99}~(3),  \pg{034502}.

\bibitem[Schneider {\em et~al.\/}(2008)Schneider, Gibson, Lagha, De~Lillo \&
  Eckhardt]{Schneider2008}
{\sc \au{Schneider, T.~M.}, \au{Gibson, J.~F.}, \au{Lagha, M.}, \au{De~Lillo,
  F.} \& \au{Eckhardt, B.}} \yr{2008}  \at{Laminar-turbulent boundary in plane
  {C}ouette flow}.  \jt{Phys. Rev. E}  \bvol{78}~(3),  \pg{037301}.

\bibitem[Schneider {\em et~al.\/}(2010)Schneider, Marinc \&
  Eckhardt]{Schneider2010}
{\sc \au{Schneider, T.~M.}, \au{Marinc, D.} \& \au{Eckhardt, B.}} \yr{2010}
  \at{Localized edge states nucleate turbulence in extended plane {C}ouette
  cells}.  \jt{J. Fluid Mech.}  \bvol{646},  \pg{441--451}.

\bibitem[Shannon(1948)]{Shannon1948}
{\sc \au{Shannon, C.~E.}} \yr{1948}  \at{A mathematical theory of
  communication}.  \jt{Bell Syst. Tech. J.}  \bvol{27}~(3),  \pg{379--423}.

\bibitem[Shi {\em et~al.\/}(2013)Shi, Avila \& Hof]{Shi2013}
{\sc \au{Shi, L.}, \au{Avila, M.} \& \au{Hof, B.}} \yr{2013}  \at{Scale
  invariance at the onset of turbulence in {C}ouette flow}.  \jt{Phys. Rev.
  Lett.}  \bvol{110}~(20),  \pg{204502}.

\bibitem[Skufca {\em et~al.\/}(2006)Skufca, Yorke \& Eckhardt]{Skufca2006}
{\sc \au{Skufca, J.~D.}, \au{Yorke, J.~A.} \& \au{Eckhardt, B.}} \yr{2006}
  \at{Edge of chaos in a parallel shear flow}.  \jt{Phys. Rev. Lett.}
  \bvol{96}~(17),  \pg{174101}.

\bibitem[Vishnampet {\em et~al.\/}(2015)Vishnampet, Bodony \&
  Freund]{Vishnampet2015}
{\sc \au{Vishnampet, R.}, \au{Bodony, D.~J.} \& \au{Freund, J.~B.}} \yr{2015}
  \at{{A practical discrete-adjoint method for high-fidelity compressible
  turbulence simulations}}.  \jt{J. Comp. Phys.}  \bvol{285},  \pg{173--192}.

\bibitem[Watanabe {\em et~al.\/}(2016)Watanabe, Iima \& Nishiura]{Watanabe2016}
{\sc \au{Watanabe, T.}, \au{Iima, M.} \& \au{Nishiura, Y.}} \yr{2016}  \at{A
  skeleton of collision dynamics: Hierarchical network structure among
  even-symmetric steady pulses in binary fluid convection}.  \jt{SIAM J. Appl.
  Dyn. Syst.}  \bvol{15}~(2),  \pg{789--806}.

\bibitem[Xi \& Graham(2012)]{Xi2012}
{\sc \au{Xi, L.} \& \au{Graham, M.~D.}} \yr{2012}  \at{Dynamics on the
  laminar-turbulent boundary and the origin of the maximum drag reduction
  asymptote}.  \jt{Phys. Rev. Lett.}  \bvol{108}~(2),  \pg{028301}.

\bibitem[Zammert \& Eckhardt(2015)]{Zammert2015}
{\sc \au{Zammert, S.} \& \au{Eckhardt, B.}} \yr{2015}  \at{Crisis bifurcations
  in plane {P}oiseuille flow}.  \jt{Phys. Rev. E}  \bvol{91}~(4),  \pg{041003}.

\end{thebibliography}

\end{document}